\documentclass[a4paper,12pt]{article}

\usepackage{setspace}
\usepackage{amsmath,amssymb,amsthm,amsfonts}
\usepackage[pdftex,xdvi]{graphicx}
\usepackage{epstopdf}
\usepackage{rotating}
\usepackage[colorlinks, citecolor=blue, urlcolor=blue, pdfborder={0 0 1}, citebordercolor={1 1 1}, pdfencoding=auto, psdextra]{hyperref}

\usepackage{enumitem}
\usepackage{geometry}
\usepackage{booktabs,longtable,multirow,colortbl,threeparttable,threeparttablex,makecell,array}
\usepackage{wrapfig}
\usepackage{float,morefloats}
\usepackage{pdflscape}
\usepackage{tabu}
\usepackage[normalem]{ulem}
\usepackage{xcolor}
\usepackage{subcaption}
\usepackage[round]{natbib}
\usepackage{comment}
\usepackage{fancyhdr}
\usepackage{xargs}

\usepackage[disable]{todonotes}
\newcommandx{\andy}[2][1=]{\todo[inline,linecolor=red,backgroundcolor=red!25,bordercolor=red,#1]{#2}}
\newcommandx{\yuan}[2][1=]{\todo[inline,linecolor=blue,backgroundcolor=blue!25,bordercolor=blue,#1]{#2}}
\newcommandx{\xinjie }[2][1=]{\todo[inline,linecolor=green,backgroundcolor=green!25,bordercolor=green,#1]{#2}}
\newcommandx{\zhentao}[2][1=]{\todo[inline,linecolor=purple,backgroundcolor=purple!25,bordercolor=purple,#1]{#2}}

\numberwithin{equation}{section}
\theoremstyle{definition}

\theoremstyle{plain}

\newtheorem{theorem}{Theorem}
\newtheorem{proposition}{Proposition}
\newtheorem{lemma}{Lemma}
\newtheorem{algorithm}{Algorithm}%[section]

\newtheorem{assumption}{Assumption}

\theoremstyle{remark}
\newtheorem{remark}{Remark}

\newcommand{\E }{\mathbb{E}}

\DeclareMathOperator{\Var}{Var}

\DeclareMathOperator{\new}{new}
\DeclareMathOperator{\tr}{tr}

\DeclareMathOperator{\ridge}{ridge}

\DeclareMathOperator{\Diag}{Diag}
\DeclareMathOperator{\rank}{rank}

\newcommand{\norm}[1]{\left\lVert#1\right\rVert}

\allowdisplaybreaks

\begin{document}

% \title{\Large\bf{Economic Forecasts with Noise}\thanks{The authors are grateful to Andrew Patton, Christian Hansen, Peter   Phillips, Yao Zheng, Oliver Linton, Tom Severini, Wenxin Jiang, Michael Pollmann and seminar participants at Duke, Tsinghua, Peking, McMaster for valuable comments. }}

% Economic forecasts with overfitting
% Economic forecasts in dense models
% Noise as regularization in economic forecasting
% Benign overfitting and the role of noise in economic forecasting

\title{\Large\bf{Benign Overfitting in Economic Forecasting \\ via Noise Regularization}\thanks{The authors are grateful to Tim Christensen, Joachim Freyberger, Christian Hansen, Benjamin Holcblat, Wenxin Jiang, Stefano Giglio, Oliver Linton, Alberto Martin-Utrera, Semyon Malamud, Michael McCracken, Ulrich M\"uller, Andrew Patton, Peter C.B.~Phillips, Michael Pollmann, Seth Pruitt, Tom Severini, Youngki Shin, Yao Zheng, and  seminar participants at Aalto, Duke, Tsinghua, Peking, Purdue, McMaster, Montreal, Wisconsin, UNC and conference participants at the SFS Cavalcade 2024 for valuable comments.%  Shi acknowledges the partial financial support from the Research Grants Council of Hong Kong No.~14617423.
}}

%\begin{comment}
\author{ Yuan Liao\thanks{Department of  Economics, University of Iowa, \texttt{yuan-liao@uiowa.edu}}
\and Xinjie Ma\thanks{Tippie College of Business, University of Iowa, \texttt{xinjie-ma@uiowa.edu}}
\and Andreas Neuhierl\thanks{Mitch Daniels School of Business, Purdue University,  \texttt{aneuhier@purdue.edu}}
\and Zhentao Shi\thanks{Department of Economics, The Chinese University of Hong Kong, \texttt{zhentao.shi@cuhk.edu.hk}}
}    
%\end{comment} 

\date{} 
\maketitle

\begin{abstract}
% slightly revised by AN Apr-12
This paper studies linear overparameterized models in economic forecasting and highlights that including noise variables (regressors with no predictive power) regularizes the estimator. We consider a setting where both the outcome variable and the high-dimensional predictors are driven by a small number of latent factors, and show that the linear forecast model is dense rather than sparse. It turns out that a ridgeless regression augmented with noise predictors attains the same asymptotic forecast accuracy as an oracle with known true factors, without estimating the factors or assuming them to be strong. The gain comes from shrinkage of the eigenvalues of the design matrix, which reduces the out-of-sample variance. In contrast, perfect variable selection that removes noise variables can worsen forecasts when the number of retained predictors is comparable to the sample size. Empirically, we apply this approach to forecasting U.S. inflation, international GDP growth, and the U.S. equity risk premium, finding that noise regularization improves and stabilizes predictive performance.

\vspace{.1in}

%\noindent {\bf Key words:} machine learning,  factor model, double descent, dense signals
\end{abstract}

\newpage

\onehalfspacing

\section{Introduction} \label{sec:intro}

Economic outcomes arise from complex, high-dimensional processes. Output growth, inflation, and asset returns are shaped by many interacting forces that are not directly observed. Economists have nonetheless sought parsimonious representations through variable selection or dimensionality reduction. But when predictors share latent factors, predictive information is diffuse: each variable carries only partial information, and no small subset captures what the full panel provides collectively. This appears to be the empirical reality. \citet{giannone2021economic} show that economic data rarely favor sparse predictive representations, and \citet{kolesar2023fragility} document the fragility of sparsity assumptions. This paper takes the denseness of economic predictive models seriously and examines its implications for the practice of forecasting. It addresses two questions: First, under what economic models can the denseness be justified? Second, what methods are appropriate, which approaches fail, and why?

In economic forecasting, observable variables are driven by a small number of latent factors representing macroeconomic, financial, or policy conditions. Because factors are unobserved, empirical practice relies on large panels of observable predictors as proxies, and estimates linear regressions as reduced-form working models. This factor structure has a fundamental implication for the reduced-form working model: the population regression coefficients are \emph{dense}. Predictive information is spread across many predictors,  and no small subset   captures  the full signal.  This  contrasts with the sparsity assumption underlying Lasso and related methods.  
Yet   when the number of predictors $p$ is comparable to the sample size $n$, the high dimensionality imposes challenges to estimating the working model for forecasts.  The standard response, namely variable selection by Lasso or dimension reduction by PCA \citep{stock2002macroeconomic, BN02}, rests on the belief that eliminating noise improves out-of-sample accuracy. This belief can be misleading. Reliable factor estimation requires sufficiently strong eigenvalue separation. When factors are weak, or when many predictors dilute factor strength across the full panel, determining the number of factors and recovering their span poses significant challenges \citep{onatski2012asymptotics, carrasco2016sample}.

Our paper develops a new approach that avoids estimating factors, whether strong or weak, while achieving the same first order optimality as if the true factors were used.  We show that when predictors substantially outnumber observations, the design matrix self regularizes.
In this regime, the Ridgeless estimator, that is, the minimum norm least squares solution computed via the Moore–Penrose pseudoinverse, benefits from  implicit regularization. Many predictors collectively dilute each variable’s influence on the fitted values, driving out-of-sample variance toward zero, a manifestation known as \textit{benign overfitting} \citep{belkin2019reconciling,mei2019generalization,hastie2022surprises}.  Variable selection, however, disrupts this mechanism.  Even perfect selection keeps the predictor-to-observation ratio near one, where self-regularization fails.  The failure is not about omitted signal, but a consequence of insufficient predictor dimension. The empirical relevance is immediate: for example, the FRED-MD dataset offers roughly 130 macroeconomic predictors for rolling estimation windows of approximately 120 monthly observations \citep{mccracken2016fred}, so the dimension to sample size ratio is close to one, a regime in which the out of sample variance becomes very large.

An appropriate response is to expand the feature space rather than contract it.   We propose a simple forecasting approach, which we call “noise-regularization”: applying the Ridgeless regression while retaining uninformative predictors. When economic theory does not provide a sufficiently rich set of informative predictors, we artificially augment the predictor set  with randomly generated noise variables independent of the data.  The key insight is that benign overfitting operates through regularization of the eigenvalue spectrum rather than predictor informativeness. Even though noise variables contain no signal, they inflate the eigenvalues of the design matrix and reduce out-of-sample variance. Because the shrinkage function is concave, Jensen’s inequality implies that this random shrinkage yields a smaller shrinkage bias in finite samples.

We formalize these insights in four results in order. Let $p_0$ denote the number of informative predictors that load on the latent factors.  We consider a setting in which a large number of economic variables, including the forecast target, are driven by a few latent factors. Our first result establishes a \emph{dense representation} of the forecasting working model (Theorem~\ref{th1}):   the population regression coefficients are dense, with their squared norm decaying as the number of important regressors   increases.  
As dimension grows,  the working model forecast converges to the oracle prediction --- the same   asymptotic mean squared error (MSE) obtained using the true factors.  The second result quantifies the \emph{cost of perfect variable selection} (Proposition~\ref{th:3ols}).  Even if an oracle correctly identifies the $p_0$ informative predictors, applying the Ridgeless regression   to them alone fails to achieve first-order optimality when $p_0/n$ converges to a positive constant. Post-Lasso, OLS on the model with perfect variable selection, or interpolation using only informative predictors all remain suboptimal.

While we acknowledge that variable selection is imperative for structural interpretation or causal inference, in this paper forecast accuracy is our focus. The third result establishes \emph{benign overfitting} (Theorems~\ref{th2} and~\ref{th4denoise}).  
The noise-regularization achieves the oracle MSE asymptotically, without estimating the factors or their number,  accommodating weak factors  and temporal dependence. When noise predictors are added deliberately and   forecasts are averaged over repeated noise draws via Rao--Blackwell, the variance rate improves  strictly.  Because    factor  estimation is bypassed, the required  factor strength conditions are mild,  and the results apply to both cross-sectional and time-series settings.

Finally, the gains from noise regularization require expanding the predictor set with genuinely exogenous variables; otherwise, overparameterization does not universally produce benign overfitting. In particular, for  the autoregressive model AR$(p)$,  adding more lagged predictors does not deliver similar gains. In the fourth result,  we   establish an AR$(p)$ \emph{symmetry property} (Proposition~\ref{th:ar}): for a time series of length $T$, the Ridgeless forecast from   AR$(p)$  numerically coincides   with that from  AR$(T-p)$  for every lag order $p$. Increasing the lag order to the overfitting regime is therefore equivalent to fitting a shorter autoregression. The intuition is that  each additional lag reduces  the effective sample by one observation, and thus lag expansion fails to induce eigenvalue inflation.

Together, these four results identify a unified mechanism: benign overfitting in dense, factor-driven forecasting environments is achievable by expanding the feature space, not by selecting within it.
Our analysis provides an economic structure for benign overfitting and extends the double descent literature \citep{belkin2019reconciling, bartlett2020benign, hastie2022surprises, lee2023mean} to factor-driven environments.  We show that the dense coefficient vector in the reduced-form working model arises naturally from the oracle factor structure, rather than being imposed as a technical condition. We further demonstrate that the second descent in forecast MSE reflects eigenvalue shrinkage, which can be induced by adding noise variables, a distinction particularly relevant  in economic studies when $p_0/n$ is not large. Finally, whereas existing analyses typically assume bounded eigenvalues, factor models feature a few rapidly diverging spikes, requiring new technical arguments.\footnote{In a contemporaneous study, \citet{meng2025estimation} analyze double descent in portfolio allocation with spiked eigenvalues. Among economic applications of overparameterization, \citet{kelly2022virtue} and \citet{didisheim2023complexity} document the benefits of rich nonlinear models.}

Empirically, we apply our approach to three settings: monthly U.S.\ inflation forecasting using FRED-MD \citep{mccracken2016fred}, cross-country GDP growth forecasting using \citet{barro1994sources}, and the U.S.\ equity risk premium forecasting  \citep{Goyal2008comprehensive}. In each application, we implement the proposed noise-regularization forecast by augmenting the predictor set with artificially generated noise variables. Across all settings and forecast horizons, predictive accuracy is stable or improves as noise predictors are added. For the U.S.\ inflation,  noise regularization achieves the lowest predictive MSE among all methods considered. For cross-country GDP growth, the method stabilizes at an MSE near 1.0, compared with 20.24 using the original 60 predictors alone, and remains robust over a wide range of noise counts. For the equity premium, where competing methods yield negative out-of-sample $R^2$, noise-regularization achieves a positive $R^2$ of approximately 1.2\%. 
It turns out that noise-regularization outperforms Ridge, Lasso, and PCA across these applications, and exhibits substantially lower sensitivity to the tuning parameter.

%%%%%%%%%%%%%%%%%%%%%%%%%%%%%%%%%%%%%%%%%%
We adopt the following notation. Let $\norm{A}$ denote the $\ell_2$ norm if $A$ is a vector, or the  operator norm if $A $ is a  matrix. For two positive sequences $a_{p,n}$ and $b_{p,n}$, we denote $a_{p,n} \ll b_{p,n}$ (or $b_{p,n} \gg a_{p,n}$) if $a_{p,n} =o(b_{p,n})$, and denote $a_{p,n} \asymp b_{p,n}$ if $a_{p,n} = O(b_{p,n})$ and $b_{p,n} = O(a_{p,n})$.

%----------------------------------------------------------------------------------------%
% OLD INTRODUCTION
%----------------------------------------------------------------------------------------%

%----------------------------------------------------------------------------------------%
% NEW SECTION II
%----------------------------------------------------------------------------------------%

\section{Model and dense representation} \label{sec:model}
\subsection{Factor-driven forecasting}\label{subsec:oracle-model}

Our objective is to forecast a scalar outcome $y_t$, driven by $K$ latent factors $f_t$ in the true data generating process
\begin{equation}\label{eq2.1}
y_t= \rho'f_t+\epsilon_{y,t},\quad \E(\epsilon_{y,t}| f_t ) = 0, \quad (\text{true DGP})
\end{equation}
where $\rho$ is a $K \times 1$ vector of factor loadings on the outcome.
The model admits an intercept by setting the first component of $f_t$ to unity.
In addition to the target, the economist observes a $p$-dimensional vector of predictors
$x_t= (x_{1,t},\ldots,x_{p,t})'$ sharing the latent factors through
\begin{equation}\label{eq2.2}
x_{t} = \Lambda f_t+ u_{t},\quad \mathbb E (u_{t}|f_t, \epsilon_{y,t})=0,
\end{equation}
where $\Lambda = (\lambda_1, \ldots, \lambda_p)'$ is a $p\times K$ matrix of factor loadings and $u_t = (u_{1t}, \ldots, u_{pt})'$ is an idiosyncratic error.
The mean independence condition $\mathbb E (u_{t}|f_t, \epsilon_{y,t})=0$ ensures that all predictive power of $x_t$ for $y_t$ operates through the common factors; Remark~\ref{rem:zero-correlation} below discusses why this condition is central to the dense forecasting setting.

We allow both time series and cross-sectional forecasts.  In the time series context, the primitive processes $(f_t, \epsilon_{y,t}, u_t)$ are stationary over $t=1,\ldots,n$, and 
 $f_t$ may be dynamic, containing finitely many lags of a baseline factor vector as in \citet{SW02}; for cross-sectional applications, $t$ indexes individuals rather than time periods.

Not all predictors need be informative.
We partition $x_t$ into components that load on and off the factors:
\begin{equation}\label{eq:Xt}
x_t =  \begin{pmatrix}
x_{I,t}\\
x_{N,t} 
\end{pmatrix} =  \begin{pmatrix}
\Lambda_I\\
0
\end{pmatrix} f_t + u_t,\quad \begin{cases}
\text{informative ones:}&  \quad \dim(x_{I,t}) = p_0\\
\text{noise:}&  \quad \dim(x_{N,t}) = p-p_0,
\end{cases}
\end{equation}
where $\Lambda_I$ is a $p_0\times K$ matrix with all rows nonzero.
The ordering is without loss of generality.
When $p - p_0$ is large relative to $p_0$, the factors are diluted across the full predictor set and become \emph{weak} in the sense of the factor-strength literature, even if they are strong within the informative block alone.
Our methodology treats the identities of $x_{I,t}$ and $x_{N,t}$ as potentially unknown and does not require variable selection to screen off the noise.

Two empirical scenarios motivate (\ref{eq:Xt}).

\medskip
\noindent \textbf{Scenario~I: Unknown noise identities.}\;
Economists routinely assemble large predictor sets such as macroeconomic indicators, financial variables, survey data that are \emph{potentially} correlated with the target, but economic theory may not pinpoint which variables carry the factor signal. Many collected predictors may be pure noise, diluting factor strength across the full set. The partition in \eqref{eq:Xt} is unknown to the forecaster.

\medskip
\noindent \textbf{Scenario~II: Intentionally added noise.}\;
Suppose, instead, that the economist has \emph{a priori} knowledge that all $p_0$ collected predictors are informative, but $p_0$ is not much larger than $n$ (formally, $p_0/n \to \gamma_0 \in (0,\infty)$). Traditional wisdom would retain only these predictors, but Proposition~\ref{th:3ols} below shows that this regime is insufficient for first-order forecast optimality, regardless of how accurately the informative variables are identified. A central finding of this paper is that deliberately augmenting the predictor set with pure noise variables, i.e.~making $p$ much larger than $n$ improves forecast accuracy through eigenvalue regularization.

\medskip

\begin{remark}\label{rem:zero-correlation}
The mean independence condition in \eqref{eq2.2} rules out direct predictability through the idiosyncratic errors, so that all forecasting power operates through the common factors as in \citet{forni1998let} and \citet{de2008forecasting}.
If instead $\epsilon_{y,t}$ is correlated with $u_t$, predictability then arises from both factors and idiosyncratic components, and the regression coefficient may be sparse: in the extreme case $y_t = x_{1,t}$, the coefficient vector is $\beta = (1,0,\ldots,0)'$.
We exclude this channel so that predictive information is spread across many variables.
\end{remark}

Now that we have set up the data generating process, we proceed to the reduced-form procedure of supervised learning. 

\subsection{Dense reduced-form representation}\label{subsec:dense}
Given the data $(y_t, x_t)_{t=1}^n$, the linear working model
\begin{equation} \label{eq3.2}
y_t =  x_t'\beta + e_t, \quad \mathbb E(x_t e_t)=0, \quad (\text{working model})
\end{equation}
with $\beta = \mathbb E(x_t x_t')^{-1} \mathbb E(x_t y_t)$, is the standard reduced-form approach to forecasting with many predictors. Economic theory implies that many outcomes depend on low-dimensional state variables (consumption growth, productivity, or risk factors) that are not directly observed \citep{merton1973intertemporal, lucas1978asset}. While no single predictor perfectly proxies the latent factors, their linear combination provides a close approximation. This suggests that the coefficient vector $\beta$ should not be sparse but dense: many small nonzero entries, rather than a few large ones.

Theorem~\ref{th1} formalizes this intuition. To state it, we introduce notation and the asymptotic regime. Let $\sigma_j(\cdot)$ denote the $j$th largest singular value of a matrix, and let $\sigma_{\min}(\cdot)$ and $\sigma_{\max}(\cdot)$ denote the smallest and largest nonzero singular values, respectively. The factor strength is indexed by a sequence $\psi_{p,n} \to \infty$.
Our asymptotic framework requires $p, p_0, \psi_{p,n} \to \infty$ as $n \to \infty$, with $p_0 \asymp n$ and $n = O(p)$.

\begin{assumption}
\label{ass:Xt}
\begin{enumerate}[label=(\roman*)]
 \item There is a sequence $\psi_{p,n}\to\infty$ such that     
    $
    \sigma_1\left( \Lambda_I'\Lambda_I\right)\asymp \sigma_K\left( \Lambda_I'\Lambda_I\right)\asymp \psi_{p,n} =O(p_0).
    $
\item  There exist finite constants $c_u , C_u \in (0,\infty)$ such that
$c_u \leq\sigma_{\min}(\Sigma_{u})\leq\sigma_{\max}(\Sigma_{u})\leq C_u$, where $\Sigma_u  = \mathbb E (u_t u_t')$.
\item $\|\rho\|$ is bounded. 
\end{enumerate}
\end{assumption}

Condition~(i) requires that all $K$ eigenvalues of $\Lambda_I'\Lambda_I$ grow at a common rate $\psi_{p,n}$. The homogeneous strength   among all factors is imposed for the ease of presentation, and can be extended to allow heterogeneous strength among factors. Crucially, we allow $\psi_{p,n}/p_0 \to 0$: the factors may be weak even within the informative block. The divergence of $\psi_{p,n}$ is necessary to separate the factor signal from the idiosyncratic component. Conditions~(ii) and~(iii) are standard. 

We now characterize the coefficient vector $\beta$ and the residual variance of the working model. Let $\sigma^2_{\epsilon} = \mathbb E(\epsilon^2_{y,t})$ denote the oracle forecast variance under the true factors, and $\sigma^2_{e} = \mathbb E(e^2_{t})$ the variance of the residual in \eqref{eq3.2}.

\begin{theorem}\label{th1} 
Suppose that $(x_t, y_t)$ are generated by models \eqref{eq2.1} and \eqref{eq2.2}, and that Assumption~\ref{ass:Xt} holds.
Then the coefficient of the working model \eqref{eq3.2} satisfies:
\begin{enumerate}[label=(\roman*)]
\item  $    \|\beta\|^2  = O(\psi_{p,n}^{-1}) =o(1)    $;
\item $ \sigma^2_{e} - \sigma^2_{\epsilon}  =O(\psi_{p,n}^{-1}) =o(1)$. 
\end{enumerate}
\end{theorem}

Result~(i) establishes that the working model is dense: as factor strength grows, the $\ell_2$-norm of $\beta$ decays to zero. This contrasts with sparse models, in which at least one coefficient remains bounded away from zero regardless of dimension. 
The conventional sparse view assumes that only a small subset of predictors carries predictive power. However, this perspective has been challenged by empirical evidence, including \cite{giannone2021economic} and many related studies. Perhaps a more convincing interpretation is that predictors are not collected arbitrarily, but are selected under the guidance of economic theory and empirical experience. As a result, many predictors are informative because they share exposure to latent economic factors. When such predictors are used directly in a reduced-form regression, each variable carries a small amount of signal through these common factors. 
As the cross-sectional dimension grows, this signal is distributed across many regressors, leading to a dense representation where $\|\beta\|\to 0$. Result~(i) formalizes this intuition and is closely aligned with the empirical findings in \cite{giannone2021economic}.

Result~(ii) quantifies the gap between the working model \eqref{eq3.2} and the oracle one (\ref{eq2.1}): the residual variances $\sigma_e^2$ and $\sigma_\epsilon^2$ coincide as $\psi_{p,n} \to \infty$.
In principle, the economist can achieve the same asymptotic forecast accuracy using the observed predictors $x_t$ directly, without estimating the latent factors or specifying their number.

%%%%%%%%%%%%%%%%

\subsection{Forecast  under variable selection}\label{subsec:variable-selection}
Given the dense nature of the working model established in Theorem~\ref{th1}, does dimension reduction or noise removal remain desirable? Consider a hypothetical oracle that only the first $p_0$ predictors load on the latent factors while the remaining $p -p_0$ are pure noise,
and thus the economist retains only the informative predictors and estimates the following ``ideal" model:
\begin{equation}\label{model:ols}
y_t = x_{I,t}' \beta_I + e_t, 
\qquad 
\dim(x_{I,t}) = p_0.
\end{equation}
Keeping $x_{I,t}$ and dropping the noise appears to be the natural course of action.
The objective is to forecast out-of-sample data $y_{\new}$ satisfying: 
\[
y_{\new} = \rho' f_{\new} + \epsilon_{y,\new}, 
\qquad
x_{I,\new} = \Lambda_I f_{\new} + u_{\new},
\]
and the in-sample and out-of-sample environment is stable so that $\sigma_\epsilon^2 = \mathbb E(y_t - \rho'f_t)^2 = \mathbb E(y_{\new} - \rho'f_{\new})^2$.

We focus on a simple tuning-free dense forecasting method, the \emph{Ridgeless regression} 
\begin{equation}\label{ols}
\widetilde\beta_I= \left(\sum_{t=1}^nx_{I,t}x_{I,t}'\right)^+\sum_{t=1}^nx_{I,t}y_t = (X_I' X_I)^{+} X_I' y,
\end{equation}
to estimate the dense coefficient in  (\ref{model:ols}),
where $\left( \cdot \right)^+$ denotes the Moore–Penrose pseudoinverse,
$X_I = (x'_{I,1},\ldots,x'_{I,n})'$, and  $y = (y_1,\ldots,y_n)'$.
This estimator coincides with OLS when $p_0 < n$ and interpolates the data when $p_0 \ge n$  (meaning $y_t= x_{I,t}'\widetilde\beta_I$ for all in-sample data). 
It turns out that the natural forecast $$\widetilde{y}_{\new}^I = x_{I,\new}' \widetilde{\beta}_I$$
is not a good practice, as shown in the following proposition.

\begin{proposition}[Informative predictors only]\label{th:3ols}

If Assumption~\ref{ass:Xt} holds and  
$u_{\new}, \epsilon_{y,\new}, X_I$, and $\epsilon_y = (\epsilon_{y,t})_{t=1}^{n}$ are mutually independent, 
then \begin{eqnarray*}
  \liminf_{n,p_0 \to \infty}\,\,\mathbb E[(y_{\new}-\widetilde{y}_{\new}^I )^2|X_I]  &>& \sigma_\epsilon^2
 \end{eqnarray*}
 as $p_0/n\to \gamma_0 \in(0, \infty)$.
\end{proposition}

The assumptions in Proposition \ref{th:3ols} require the regression error terms $(\epsilon_{y,\new}, u_{\new})$ to be independent of the in-sample data; however, we allow correlations between $f_{\new}$ and the in-sample $f_t$
to accommodate dynamic factors in time series settings. 
The proposition establishes that even perfect variable selection fails to achieve first-order forecast optimality. The failure is not due to omitted signal --- all retained predictors are informative. Rather, when $p_0$ is proportional to $n$, the dimension is not sufficiently large to regularize the eigenvalues of the design matrix $X_I'X_I$,  which is crucial for reducing the out-of-sample variance.  Hence Ridgeless does not benefit from benign overfitting.

This result should be interpreted  broadly as an extension of the classical post-variable-selection phenomenon to the overparameterized regime.  Suppose a  dimension reduction method, such as  \citet{belloni2013least}'s Post-Lasso, identifies the informative predictors without error, and OLS is then applied to the selected set. When $p_0 < n$, the oracle Ridgeless estimator $\widetilde{\beta}_I$ coincides with OLS, making it equivalent to Post-Lasso under perfect selection. Proposition~\ref{th:3ols} illustrates that even in this idealized setting, the resulting forecast fails to achieve first-order optimality when $p_0/n \to \gamma_0 \in (0,\infty)$.

\medskip

\noindent \textbf{Symmetry of AR($p$).}\;
If eigenvalue inflation cannot be achieved by retaining the informative predictors only, one might instead seek overparameterization by expanding model complexity in a different direction. In time series forecasting, an off-the-shelf candidate is to increase the lag order in an autoregressive model. Consider a time series of length $T$, where the objective is to forecast $y_{\new} = y_{T+1}$.
Let $y_{a:b} = (y_t)_{t=a}^b$ denote the column vector of $y_t$ for periods $a$ to $b$.
Consider the AR($p$) model with no intercept, estimated by the Ridgeless regression, which is equivalent to running AR($p$) with the demeaned data.  The Ridgeless AR($p$) coefficient  and one-step-ahead forecast are
\begin{equation}\label{arp1}
\widehat \beta^{AR}_{p} = \left(\sum_{t=p+1}^{T} y_{t-p:t-1} y_{t-p:t-1}'\right)^+ \sum_{t=p+1}^{T} y_{t-p:t-1} y_{t},
\end{equation}
\begin{equation}\label{arp2}
\widehat y^{AR}_{T+1}(p) = y_{T-p+1:T}'\widehat \beta^{AR}_{p}.
\end{equation}
For AR($p$),  the effective sample size is $n=T-p$.   So  \eqref{arp1} coincides with OLS when $p < T/2$ and interpolates when $p \ge T/2$.

If benign overfitting were driven simply by increasing the number of regressors, one would expect gains from choosing $p \geq  T/2$. The following proposition, however, shows that this intuition breaks down due to a “symmetry” property of AR($p$).

\begin{proposition}[Symmetry of AR($p$)]\label{th:ar}
Let $\widehat y^{AR}_{T+1}(p)$ be the Ridgeless AR$(p)$ forecast defined in \eqref{arp1}--\eqref{arp2}. The equivalence
$$
\widehat y^{AR}_{T+1}(p) = \widehat y^{AR}_{T+1}(T-p)
$$
holds for any $p=1,\ldots,T-1$.
\end{proposition}

\begin{figure}[h!]
\caption{The symmetry property of AR$(p)$ forecast using Ridgeless.} \label{fig:ar_symmetry}
\begin{center}
    \includegraphics[width=0.7\textwidth]{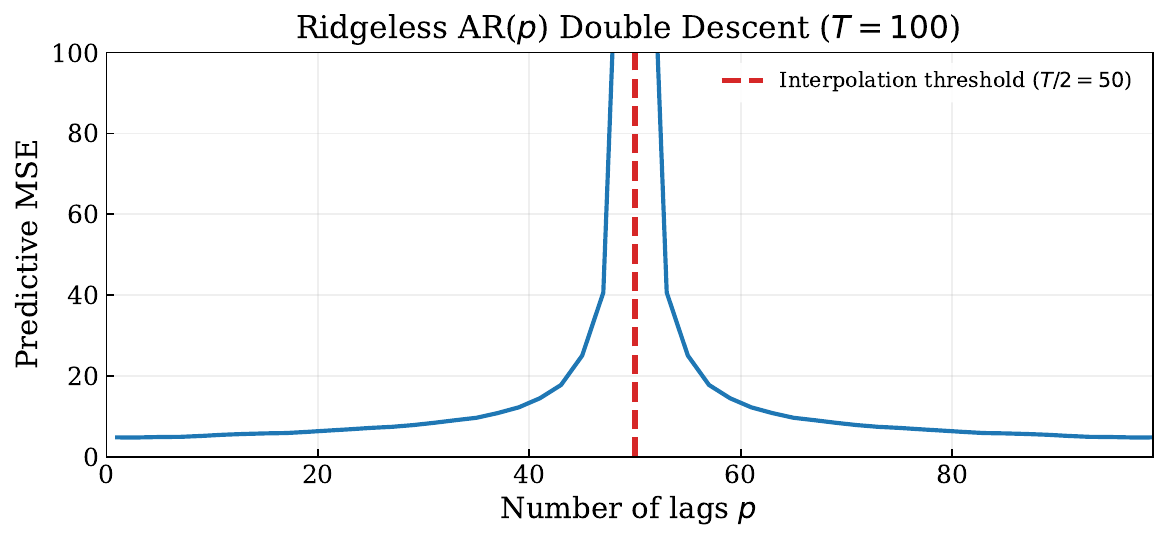}
\end{center}
\vspace{-4mm}
\begin{flushleft}
\footnotesize Notes:  The blue curve plots the predictive MSE. The time series length is $T=100$. The interpolation threshold is $T/2=50$. When $p< T/2$, the predictor is equivalent to the conventional OLS.
\end{flushleft}
\end{figure}

Proposition \ref{th:ar} reveals a  novel  symmetry property: an   overparameterized AR($p$) model with $p > T/2$ is numerically identical to an underparameterized AR($T-p$) model.   As $p$ increases, the effective sample size $n = T - p$ decreases proportionally, so the eigenvalues of the design matrix do not grow.% in a manner that induces spectral regularization. 
Lag expansion therefore does not generate the eigenvalue inflation necessary for benign overfitting. Figure~\ref{fig:ar_symmetry} illustrates this symmetry.

\medskip

Taken together, Propositions~\ref{th:3ols} and~\ref{th:ar} convey a unified message: neither oracle variable selection nor lag expansion delivers eigenvalue growth in the design matrix. %, which is needed for benign overfitting.
Under oracle selection, $p_0$ proportional to $n$ leaves eigenvalues insufficient in magnitude; under AR($p$), the effective sample size shrinks as the lag order grows.
The central lesson is that the spectrum, not informativeness alone, governs forecast accuracy in the overparameterized regime. 
This motivates the approach developed in Section~\ref{sec:noise_reg}, where we expand the predictor dimension to induce the spectral regularization.

\medskip

\noindent \textbf{Why not Lasso or PCA?}\; 
Theorem~\ref{th1} establishes that $\beta$ is dense, which immediately limits the scope of the Lasso \citep{chernozhukov2017lava, giannone2021economic}. In addition, Proposition \ref{th:3ols} covers the Post-Lasso under perfect variable selection as a special case, which also yields suboptimal forecast.   
PCA offers an alternative by estimating the factors directly, but the quality of the estimated factors depends critically on factor strength.
When many noise variables are present, the factors appear weak across the full set of predictors, and correctly estimating the number of factors is difficult in practice; see recent work by \citet{giglio2023prediction} and \citet{chao2022selecting} from the perspective of feature selection.
Furthermore, weak factors need not arise solely from noise contamination; factors can also be weak within informative predictors.  If the eigenvalues of $\Lambda_I'\Lambda_I/p_0$ decay to zero as $p_0 \to \infty$ (so that $\psi_{p,n}$ grows slowly relative to $p_0$), the factor signal is intrinsically weak and PCA continues to suffer from poor factor recovery regardless of how successfully noise variables are screened.

%----------------------------------------------------------------------------------------%
%  SECTION III
%----------------------------------------------------------------------------------------%

\section{Noise regularization} \label{sec:noise_reg}
Under the conditions of Theorem~\ref{th1}, the working model is dense, meaning that a non-negligible fraction of predictors contributes to predictive performance. In practice,  the predictor set comprises the informative variables $X_I$ ($n \times p_0$) and the noise variables $X_N$ ($n \times (p - p_0)$).
Regardless of whether the economist   knows the identities of   informative  predictors,  Proposition~\ref{th:3ols} shows that informative predictors solely are insufficient for first order forecast optimality when the number of informative predictors $p_0$ satisfies $p_0 / n \to \gamma_0 < \infty$. This is empirically relevant because stylized economic theory may provide only a few informative predictors.

This motivates our  proposed forecast approach: rather than eliminating noise variables, one should retain all available predictors and, when necessary, intentionally add pure noise variables to expand the predictor set. Importantly, our approach does not distinguish between the two scenarios described in Section \ref{sec:model}. 
Formally, let $X = [X_I, X_N]$ be the $n \times p$ matrix pooling all predictors. In the first scenario, $X_N$ includes the noise variables already present in the dataset; in the second scenario, $X_N$ consists of artificially added noise columns. In both cases we apply the Ridgeless regression on  predictors with noise:
\begin{equation}\label{rtidours}
   \widehat\beta = (X'X)^+ X'y. 
\end{equation}
We refer to (\ref{rtidours}) as a “noise regularized estimator,” regardless of whether the noise predictors arise from the dataset or are intentionally introduced by the econometrician. A central contribution of this paper is to show that high-dimensional noise variables serve as an implicit regularization device: they inflate the non-factor eigenvalues of $X'X$ (those not corresponding to the factors), thereby reducing the out-of-sample variance.  The next subsection explains this regularization mechanism in detail.

\subsection{Noise-regularization as a shrinkage mechanism}\label{sec:noiseregul}
Given that the added noise contains no information, how does it facilitate prediction? It transpires that adding noise can be viewed as a regularization method, which inflates the eigenvalues of the design matrix.   To see this, define the following $p_0\times 1$ vector-valued shrinkage function: 

$$
\mathcal A(M)=   X_I'(X_IX_I'+ M)^{-1}y,
$$
where $M$ is an $n\times n$ symmetric positive semidefinite matrix, provided that the inverse exists.  When $p>n$, the definition of pseudoinverse gives $(X'X)^+ X = X' (XX')^{-1}$.  Therefore the subvector of    the noise-regularized estimator (\ref{rtidours}) corresponding to   the informative predictors  can be written as 
 $$
 \widehat\beta_I= \iota_I(X'X)^+X'y = X_I'(XX')^{-1}y
 =\mathcal  A(X_NX_N')
 $$
where, without loss of generality, the first $p_0$ predictors are the informative ones picked out by the $p_0\times p$ selection matrix $\iota_I=(I_{p_0},0)$, and the last equality decomposes the $n\times n$ matrix $XX'= X_IX_I'+ X_NX_N'$, which is invertible when $p>n$.  As a comparison,  perfect variable selection specifies $M = 0$, and the Ridge regression with tuning parameter $\lambda$ sets $M = \lambda I_n$. The choice of $M$ governs how aggressively the eigenvalues of $X_IX_I'$ are inflated; a larger $M$ makes $(X_IX_I' + M)^{-1}$ smaller in the semidefinite order, directly controlling the out-of-sample variance.

To see the effect of shrinkage using a noise matrix,  take the singular value decomposition (SVD):
\begin{equation}
\underset{(n\times p_0)}{X_I}= \underset{(n\times s)}{U_I} \underset{(s\times s)}{D_I} \underset{(s\times p_0)}{V_I'}  ,\quad s=\min\{p_0,n\}
\end{equation}
where $D_I$ is the diagonal matrix of nonzero singular values of $X_I.$ Define 
 $$
\mathcal S(M,\Sigma)=\tr\left[V_ID_I(D_I^2+U_I' M U_I)^{-2}D_IV_I'\Sigma\right]
 $$
 
Consider, for simplicity, the case when   $x_{\new, I}$ is independent of the training data (to be relaxed in Assumption \ref{ass:e}). It follows that the out-of-sample variance is 
 $$
 \Var(x_{\new,I}'(\widehat \beta_I- \beta_I)|X)= \mathcal S(X_NX_N',\Sigma_{X_I}).
 $$
When $M=X_NX_N'$, it inflates the eigenvalues from $D_I^2$ along a non-factor direction to   $D_I^2+U_I' M U_I$.  Meanwhile, it is well-known from the random matrix theory that the nonzero eigenvalues of the noise matrix $X_NX_N'$ grow proportionally to $p$ if $ p/n \to \infty$. As a consequence, the order of the non-spiked  (the largest $(K+1)$-th to the $n$-th) eigenvalues change from $p_0$ to the order of $p.$ 
Also for factor models, we have $\Sigma_{X_I}=\Lambda_I \Sigma_f \Lambda_I'+  \Sigma_{u_I}$, which yields a further decomposition of the variance into the factor-related component and the idiosyncratic component, and for now  let us focus on the latter. This yields  
$$
 \mathcal S(X_NX_N', \Sigma_{u_I})\leq C_u\tr\left[(D_I^2+U_I'WU_I)^{-2}D_I^2\right] =  O_P\left(\frac{n+\psi_{p,n}}{p^2}\right),
 $$ which has a fast converging out-of-sample forecast variance.

In comparison, if we do not shrink the eigenvalues but  only use $X_I$   after perfect variable selection, the Ridgeless estimator (\ref{ols}) would take the form:
 $$
 \widetilde\beta_I=  (X_I'X_I)^+X_I'y =\mathcal  A(0\cdot I_{n }), 
 $$
 whose forecast variance is 
 $$
 \Var(x_{\new,I}'(\widetilde \beta_I- \beta_I)|X)= \mathcal S(0\cdot I_{n },\Sigma_{X}).
 $$
Here $M = 0\cdot I_{n }$ does not shrink the eigenvalues of $D_I^2$ and thus is of order $p_0$. Then under Assumption \ref{ass:Xt} there exists a $0 < c_u \leq \sigma_{\min}( \Sigma_{u_I}  )$ such that
$$
\mathcal S(0\cdot I_n,\Sigma_{u_I})\geq c_u \tr(D_I^{-2}) \asymp \frac{n}{p_0}.
$$
The variance does not converge when $n/p_0\nrightarrow 0$.

Inclusion of many random noise variables in essence inflates the eigenvalues to a much larger magnitude to reduce   the out-of-sample variance.  In this sense, it is similar to the Ridge regression with a large choice of tuning parameter, which also shrinks the eigenvalues.
It is therefore  important to see how the noise-regularized estimator compares with the Ridge estimator.  

The Ridge estimator, when  applied to $X_I$ with tuning parameter $\lambda$, 
takes the form 
 $$
 \widetilde\beta_{\ridge}(\lambda)=  (X_I'X_I + \lambda \cdot I_n)^{-1} X_I' y = \mathcal A(\lambda \cdot I_{n}). 
 $$
 In addition,  since $
  \mathbb  E X_NX_N' =  (p-p_0)I_n 
$,
a natural benchmark of the Ridge  tuning parameter is $\lambda= p-p_0$. 

Hence the noise-regularized estimator uses random shrinkage $ X_NX_N'$, whereas Ridge uses the deterministic shrinkage $\lambda\cdot I_n=\mathbb EX_NX_N'$. Both   estimators  introduce shrinkage bias, which will be compared heuristically.  Conditioning on the informative predictors, the estimation bias of any estimator taking the form 
$ \widehat\beta= \mathcal A(M) $  is given by
$$
\mathbb E(\widehat\beta|X_I)-\beta= \mathbb E[\mathcal A(M)|X_I]  - \beta =\mathbb E[f(M)|X_I]\beta,
$$
where 
$
  f(M)=  X_I'(X_IX_I'+M)^{-1}X_I- I_{p_0}.
$
Note that $f(M)$  is  a convex matrix-valued function, and is a negative semidefinite matrix   whenever $M$ is  positive semidefinite.  Applying the matrix-valued Jensen's inequality (e.g., \cite{tropp2015introduction}),  
$$
\mathbb E[f( X_NX_N')|X_I] \succcurlyeq f(\mathbb E( X_NX_N'|X_I)) =f(\lambda\cdot I_n), 
$$
where we write $B \succcurlyeq A$ (equivalently $A \preccurlyeq B$)  if $A-B$ is negative semidefinite.  Hence, the shrinkage effects to $\beta$ have the following relationship:
$$
\underbrace{\mathbb E[f(\lambda\cdot I_n)|X_I]}_{\text{Ridge bias}} \preccurlyeq \underbrace{\mathbb E[f(X_NX_N')|X_I]}_{\text{Noise-regularized bias}} \preccurlyeq 0\cdot I_n.
$$
Intuitively speaking, this means the shrinkage bias introduced by the random matrix $X_NX_N'$ is less than that of the deterministic matrix $\mathbb EX_NX_N' =\lambda\cdot I_n$ with $\lambda = p - p_0$, yielding a smaller finite-sample shrinkage bias than that of the corresponding Ridge.

\begin{figure}[h!]
 \begin{center}
 	\includegraphics[scale = 0.35]{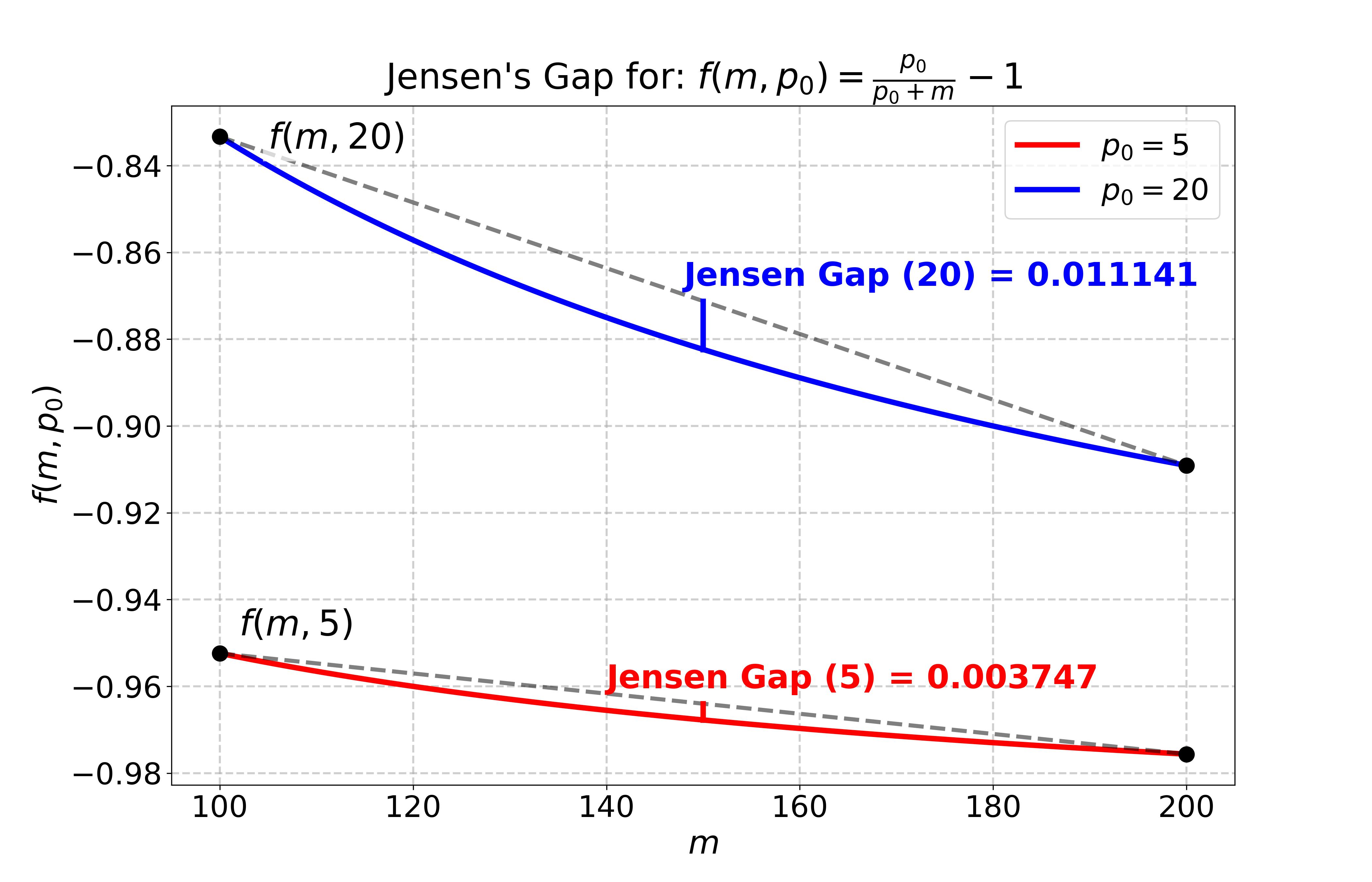}
		\caption{Jensen's Gap}\label{fig:jensen} 

\end{center}
\begin{flushleft}

\footnotesize
Notes: This plots the function $f(m, p_0)$ for two values of $p_0\in\{5,10\}$. It illustrates that the Jensen gap enlarges as $p_0$ increases.  
   
\end{flushleft}
\end{figure}

Interestingly, the bias improvement over Ridge becomes more substantial as $p_0$ increases. This can be seen by examining the  Jensen's gap, i.e., the difference between the upper and lower side of the Jensen's inequality. Consider the scalar version of $f(M)$, say $f(m, p_0)= \frac{p_0}{p_0+m}-1$. The Jensen gap is defined as
$$
\mathbb Ef(m,p_0) - f(\mathbb E m ,p_0).
$$
As shown in Figure \ref{fig:jensen}, when $m$ is much larger than $p_0$, the Jensen gap widens as $p_0$ grows. In the matrix-valued case, this corresponds to the case that when $\dim(X_N)> \dim(X_I),$ a larger  dimension  of $X_I$ yields a larger  reduction  of the shrinkage bias compared to the Ridge.

To summarize, the shrinkage representations through $\mathcal{A}(M)$ reveal how the structure of regularization differs. Perfect variable selection fixes $M = 0$ and provides no eigenvalue inflation, leaving the variance bounded away from zero. Ridge regression sets $M = \lambda\cdot I_n$ and imposes uniform, deterministic shrinkage. Noise regularization specifies $M = X_NX_N'$ and induces random shrinkage that, by the convexity of $f$, produces less bias than its Ridge counterpart while achieving the same variance reduction.

Below, we formalize this intuition of regularization, by establishing the first-order forecast optimality: The Ridgeless regularized by  sufficiently many noise achieves the same asymptotic MSE as if the true factors were directly used for prediction. We shall  present the results separately for the two scenarios.

\subsection{Scenario I: identities of noise are unknown}
\label{sec:unknown}

In the first scenario, the economist   does not know the identities of informative and noise variables. She uses  all predictors  in $X=[X_I, X_N]$,   and predicts $y_{\new}$ by
$$
\widehat y_{\new} = x_{\new}' \widehat \beta.
$$
We impose technical assumptions below to establish the benefit of forecast with noises.  Recall the data generating process in \eqref{eq2.1} and \eqref{eq2.2} shared by the training data and the test data indexed by for $t\in \{1,\ldots,n\}\cup\{\new\}$.

\begin{assumption}\label{ass:e}

\begin{enumerate}[label=(\roman*)]

\item The innovation % to $(X', x_{\new})'$ 
satisfies 
$$ (U', u_{\new})= \Sigma_u^{1/2} (W', w_{\new}), $$
where $(W', w_{\new})'$ is an $(n+1)\times p$ matrix of i.i.d.~entries with zero mean, unit variance, and finite 4th moment. 
 In addition,  $(W', w_{\new})'$ and 
 $\{ (f_t, \epsilon_{y,t})_{t \in \{1,\ldots,n\}\cup\{\new\}}  \}$ are independent.

\item Conditioning on $(F, f_{\new})$, the error terms $(\epsilon_{y,t})_{t \in \{1,\ldots,n\} \cup\{ \new\}}$ are homoskedastic, zero mean, and independent over time.

\item $\left\Vert \mathbb{E}\left[f_{\new}f_{\new}'|F \right]\right\Vert \leq C_f$ for a positive constant $C_f$.

\item The top $K$ eigenvalues of  $\Lambda_I \Sigma_f\Lambda_I'/\psi_{p,n}$  are distinct.

\end{enumerate}

\end{assumption}
 
Assumption \ref{ass:e} (i) imposes that $(U', u_{\new})$ are linear transformations of the two-way i.i.d.~matrix with a finite 4th moment,
which allows us to invoke results from the random matrix theory to reveal the behavior of   empirical eigenvalues of the idiosyncratic  sample covariance matrix $U'U/n$ of large dimensions.  
The independence between $(W', w_{\new})'$ and
$\{ (f_t, \epsilon_{y,t})_{t \in \{1,\ldots,n\}\cup\{\new\}}  \}$ strengthens the mean independence in \eqref{eq2.2}
 to simplify the computation of the variance of the estimator. 
Condition (ii) regulates the unpredictable error in the target variable. Though it is possible to extend to  {martingale difference sequences}, 
we maintain the independence to keep it in line with the assumption on $u_t$.
Condition (iii) is a technical assumption that bounds the conditional variance of the new factor realizations, which accommodates the time series case when $f_{\new}$ is correlated with $F$.  Finally, Condition (iv) ensures the uniqueness of the eigenvectors associated with the distinct top $K$ eigenvalues.

\medskip
\begin{remark}
Our method allows for both cross-sectional and time series forecasts. In the context of time series forecasting, we emphasize that 
both $y_t$ and $x_t$ may exhibit serial dependence, 
through the dynamics in factors.   What is ruled out by Assumption \ref{ass:e} (i) and (ii) is the serial dependence of the idiosyncratic errors $(\epsilon_{y,t},u_t)$ over time,   to simplify the arguments. 
This condition resembles what \citet[p.621]{hirano2017forecasting} impose on the regression error ($e_t$ in our notation) 
when they calculate the prediction risk, 
and \citet[pp.1016--1017]{doz2012quasi} in quasi-maximum likelihood estimation.
\end{remark}

\medskip

Conditioning on $X$, the predictive MSE is 
$$
\mathbb E[(y_{new}-\widehat y_{\new})^2|X]
= \sigma_\epsilon^2+ \mathbb E\left[(\rho'f_{\new}-  \widehat {y}_{\new} )^2|X\right]
$$
The first term   $\sigma_\epsilon^2  $ is the minimum forecast MSE under the true latent factors. The second term, depending on the estimator, can be decomposed as
\begin{align} 
\label{eq:decomp1}
\rho'f_{\new}-\widehat{y}_{\new}
 =\left(\rho'f_{\new}-x_{\new}'\beta\right)-x_{\new}'(\widehat{\beta}-\beta),
\end{align}
where $ \rho'f_{\new}-x_{\new}'\beta $
is the approximation error of the working linear model \eqref{eq3.2}.
The estimation error  $x_{\new}'(\widehat{\beta}-\beta)$   can be further
decomposed into 
\begin{eqnarray}\label{eq:decomp2}
    x_{\new}'(\widehat{\beta}-\beta)
&=&x_{\new}'\delta_\beta+x_{\new}'B_{X}\epsilon_{y},\\
  \text{where } \delta_\beta &=&\underbrace{ A_{X}\beta}_{\text{projection}}+ \underbrace{ B_{X}\left(F(\rho-\Lambda'\beta)-U\beta\right)}_{\text{working model approx.~error}}
\end{eqnarray}
with $A_X= (X'X)^+X'X-I_p$ and  $B_X= (X'X)^{+}X'$. 
The term $x_{\new}' \delta_\beta$ can be viewed as the bias (conditioning on $(U,F$)).
The other term $x_{\new}'B_{X}\epsilon_{y}$  constitutes the  variance due to the innovation $\epsilon_y$, 
and it is independent of $\beta$. Since $$\mathbb E[[ x_{\new}'(\widehat{\beta}-\beta)]^2|X]\leq 2\mathbb E[( x_{\new}'\delta_\beta)^2|X]+ 2\mathbb E[( x_{\new}'B_X\epsilon_y)^2|X],$$ the  following theorem presents the convergence of each component on the right-hand side.

\begin{theorem}\label{th2}
Suppose   Assumptions \ref{ass:Xt}  and \ref{ass:e} hold. If $p=o(\psi_{p,n}n)$ and $n=o(p)$, then:

\begin{enumerate}[label=(\roman*)]

\item The components of the forecast error is:
 \begin{eqnarray*}  
   \mathbb E [(x_{\new}'\delta_\beta)^2 |X ] &=& O_P\left(\frac{p}{\psi_{p,n}n}  + \frac{n}{p} \right) ,\quad \mathrm{(bias)} \cr
\mathbb E [(x_{\new}'B_{X} \epsilon_y )^2 |X ]&=& O_P\left(\frac{1}{n}+\frac{n}{p}\right), \quad \mathrm{(variance)}.
 \end{eqnarray*}

\item As $n,p\to \infty$, 
$$
\mathbb E[(y_{\new}-\widehat y_{\new})^2|X] \stackrel{P}{\to}\sigma_{\epsilon}^2.
% \Var(\epsilon_{y,\new}|X).
$$

\end{enumerate}
\end{theorem}

Theorem \ref{th2} implies that keeping the noise variables and applying the Ridgeless forecast, we can achieve the oracle predictive MSE as if  the latent factors were   used directly in forecasting. This result allows factors to be relatively weak among the informative predictors, i.e., $\psi_{p,n}=o(p_0)$ in Assumption \ref{ass:Xt} (i).

Contrary to conventional wisdom suggesting that  the variance is amplified under overfitting,  here the variance diminishes as $p\gg n$. 
The reduction of variance requires no condition on the predictability, i.e., it does not matter whether the predictors are mostly noise or informative.  This is the phenomenon as documented as  ``benign-overfitting" in the literature.  But different from the literature, the overfitting is mainly caused by high-dimensional uninformative predictors $X_N$. In the presence of many noise variables the out-of-sample variance decays as $p\to\infty$.

\subsection{Scenario II: intentionally added noise}
\label{sec:known}

When economic theory and empirical evidence guide data collection, regressors are likely to be informative, but the number of available predictors $p_0$ may not be very large, say, just proportional to $n$. 
This occurs, for instance, in diffusion index forecasts using FRED-MD as mentioned in Introduction.

We recommend adding noise to intentionally increase the overall dimensionality.  The inclusion of noise predictors inflates eigenvalues of the design matrix  and thus reduces the out-of-sample variance.  In the meantime,  inspired by the Rao–Blackwellization for risk reduction \citep{hirano2017forecasting}, the impact of randomness can be smoothed out by taking the average of repeated Ridgeless predictions.

Recall that $(X_I, y)$ denotes the in-sample data in this scenario, and the corresponding out-of-sample predictor is $x_{\new, I}$, where both $X_I$ and $x_{\new, I}$ are informative predictors.

\begin{algorithm} Noise-regularized forecast. 
\begin{description}
    \item[Step 1 (adding noise):] Generate $n\times (p-p_0)$ i.i.d.~$N(0,1)$ matrix as the noise $X_N$. Merge with $X_I$ to get $X=(X_I, X_N)$, and obtain the Ridgeless estimator
$$
\widehat \beta= (X'X)^+X'y.
$$
 Let $\widehat\beta_I$ be the subvector of $\widehat \beta$ corresponding to $X_I$ and predict
 $\widehat   y_{\new, I}:= x_{\new, I}' \widehat\beta_I.$ 
 
 \item[Step 2 (denoising):] 
 Repeat Step 1 to obtain $ \widehat \beta_{I}^b$ and then 
 $\widehat{y}^b_{\new} = x_{\new, I}' \widehat \beta_{I}^b$ for an index $b = 1,\ldots, B$. Average over the $B$ replications to get a forecast
 $$
\widehat y_{\new}^* =  B^{-1}\sum_{b=1}^B \widehat y_{\new, I}^b = x_{\new, I}' \widehat{\beta}_I^*
 $$
where $\widehat{\beta}_I^* = B^{-1}\sum_{b=1}^B \widehat{\beta}_{I}^b$.
\end{description}
 
\end{algorithm}

Step 1 creates artificial pure noise $X_N$ to be merged with $X_I$, and then uses only the subvector $\widehat\beta_I$ for prediction; the forecast depends on the realization of $X_N$. Step 2 averages out the randomness caused by $X_N$ through the $B$ repetitions. This estimator approximates
\begin{equation}\label{bl35}
    \mathbb E(\widehat   y_{\new, I}|X_I, y, x_{\new, I})
\end{equation}
with a sufficiently large $B$. This estimator can be connected to  Bayesian model average: it arises as an equally weighted  average   forecasts (conditioning on the data) from  $B$  submodels $\widehat y^b_{\new, I}$, where each submodel is implied by the drawn random noise,  whose posterior probability is $1/B$.

As a model averaging estimator, it reduces the mean squared error of   submodels.  With a slight abuse of notation, we shall use  $\widehat y_{\new}^* $ to denote (\ref{bl35}); it is a function of the observed data only. It follows that
 $$ x_{\new}'\beta-\widehat  y_{\new}^*= \mathbb E(x_{\new}'\beta- \widehat y_{\new, I} | X_I, y,x_{\new, I}),$$
and the law of iterated expectations gives
\begin{eqnarray}
    \mathbb E[(x_{\new}'\beta- \widehat  y_{\new}^*)^2|X_{I}]
  &  = &    \mathbb E\left\{\left[\mathbb E(x_{\new}'\beta- \widehat y_{\new, I }  |X_I, y,x_{\new, I})\right]^2\bigg{|}X_I\right\}\cr 
  &\leq & \mathbb E\left\{ \mathbb E\left[(x_{\new}'\beta- \widehat y_{\new,I } )^2|X_I, y,x_{\new, I}\right]\bigg{|}X_I\right\}\cr
  &=&  \mathbb E\left[(x_{\new}'\beta-\widehat y_{\new,I } )^2|X_I\right],
\end{eqnarray}
where the inequality follows by the celebrated Rao-Blackwell theorem. It shows that $\widehat  y_{\new}^*$ enjoys a smaller predictive MSE than that of a single realization $\widehat y_{\new, I}$. 

In this scenario we define $x_{\new} = (x'_{\new, I}, 0')'$ and $\beta = (\beta'_I,0')'$. Then 
$$\mathbb E[ ( \widehat y_{\new}^*-x_{\new}'\beta)^2|X_I]\leq 2\mathbb E[( x_{\new}'  \delta_\beta^*)^2|X_I]+ 2\mathbb E[( x_{\new}'B^*_X\epsilon_y)^2|X_I]$$
where
\begin{align}\label{eq:decomp2-star}
  \delta_\beta^*  =  A^*_{X}\beta +  B^*_{X}\left(F(\rho-\Lambda'\beta)-U\beta\right).
\end{align}
with $A_X^* = \mathbb E[ (X'X)^+ X'X | X_I]- I_p$ and $B_X^* = \mathbb E[ (X'X)^+ X |X_I] $.
We have the following result for the noise-regularized forecast.

\begin{theorem}\label{th4denoise}
Under the assumptions in Theorem \ref{th2} and $n p_0 = o(p^2)$, we have 
\begin{enumerate}[label=(\roman*)]
\item The forecast squared bias and variance:
 \begin{eqnarray*}
  \mathbb E [(x_{\new}' \delta_\beta^*)^2 |X_I ] &=& O_P\left(\frac{p}{\psi_{p,n}n}\right),\quad \mathrm{(bias)} \cr
\mathbb E [ (x_{\new}'B^*_{X}\epsilon_{y})^2 |X_I]&=& O_P\left(\frac{1}{n}+\frac{np_0}{p^2}\right). \quad \mathrm{(variance)}.
 \end{eqnarray*}

\item As $n,p_0, p\to \infty$, 
$$
\mathbb E[(y_{\new}-\widehat y^*_{\new})^2|X_I] \stackrel{P}{\to} \sigma_{\epsilon}^2.
$$

\end{enumerate}
\end{theorem}

This theorem shows that adding the  noise enables achieving the oracle forecast MSE as if the true factors were used.   In addition, the rate of convergence of the forecast variance is faster than that of Scenario I when $p=o(n^2)$, as a result of   the Rao–Blackwellization step.

\medskip

\begin{figure}[h!]
 	\hspace{-3em} 
    \begin{center}
    \includegraphics[width=0.48\textwidth]{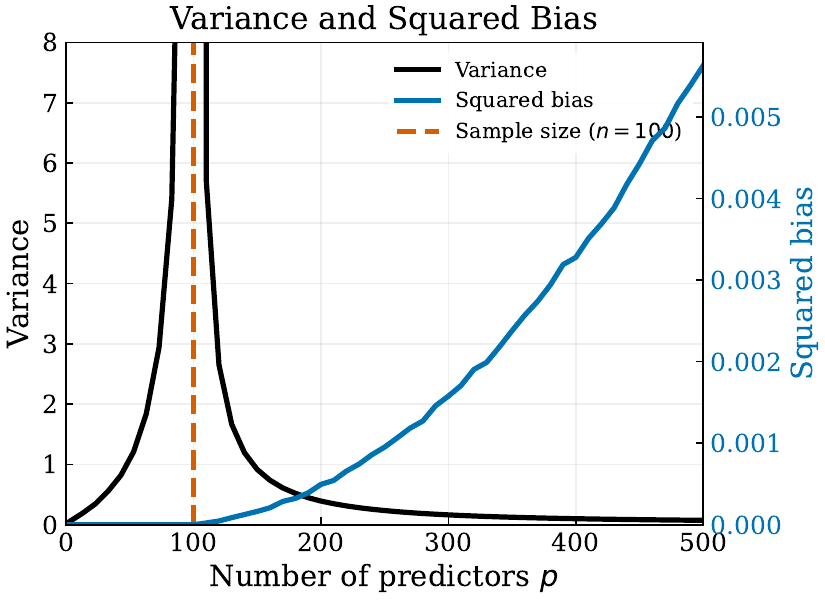}
    \includegraphics[width=0.48\textwidth]{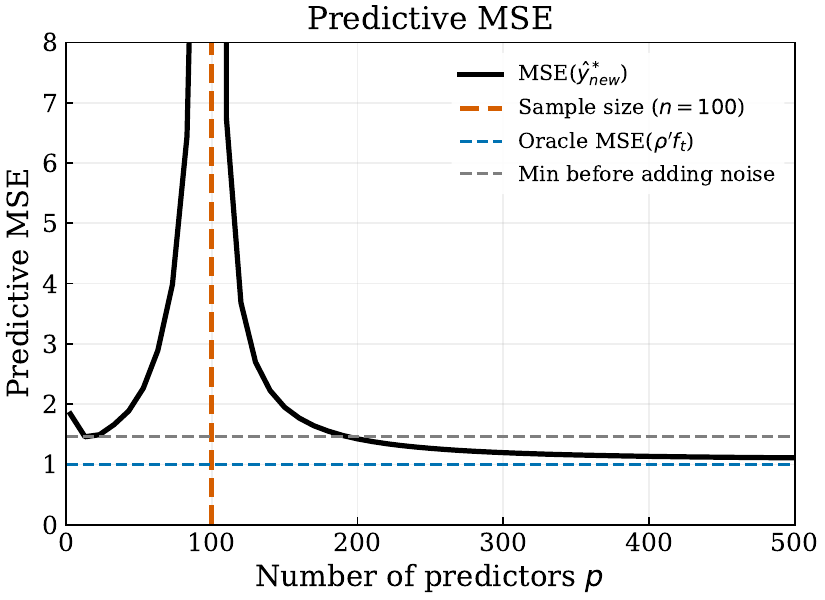}
		\caption{Bias-Variance Tradeoff } \label{fig:estK}
    \end{center}
\begin{flushleft}
    \footnotesize Notes: Theoretical predictive variance and squared bias (left panel) and MSE (right panel), averaged over 500 replications. The horizontal axis is the number of predictors increasing from 3 to 500, and we fix $n=100$.  The first 
  $p_0=\min\{p,n\}$ are   informative predictors, generated using a 3-factor model of  strong factors. The remaining $p-p_0$ are i.i.d.~Gaussian noise. The vertical dashed line is   where $p$ equals $n$, and the horizontal  dashed line on the   right panel refers to MSE$(\rho'f_t)=\Var(\epsilon_{y,t})$, the  oracle predictive MSE.  
\end{flushleft}
\end{figure}

To provide a quick demonstration, we show in 
Figure \ref{fig:estK} the theoretical curves of bias-variance (left panel) and predictive MSE $\mathbb E[(y_{new}-\widehat y^*_{\new})^2|X_I]$ (right panel)  in a 3-factor model. Here the first  $p_0=\min\{p,n\}$ predictors are informative, while the remaining $p-p_0$ predictors are i.i.d.~$N(0,1)$ white noise variables.  As is clearly illustrated, the variance monotonically increases as $p$ increases even though the first $p_0$ added predictors are all informative, and peaks at $p=n$ where the in-sample data are perfectly interpolated. Meanwhile, after $p>n$, the added predictors are pure noise, and the variance starts to decay. This is consistent with our theory: as $p\to\infty$, the variance  continues to decrease until  the $1/n$ term becomes  dominant. In addition, the squared bias remains zero until  $p=n$.
Though after passing this threshold it starts to increase, it is in a much smaller magnitude than that of the variance.  This is also consistent with the theory. 
The depicted bias does not diminish  because here we fix $\psi_{p,n}\asymp  n$ at $n=100$ while we vary $p$.

Overall, the predictive MSE (right panel) illustrates the double descent phenomenon, where the first descent occurs   before $p<20$, due to the decay of the gap $\sigma_e^2- \sigma_{\epsilon}^2$. The second descent occurs after  $p>n$, due to the decay of variance, and eventually the MSE approaches the oracle MSE as if the latent factors were used for prediction. While double descent has been documented in the recent statistical literature, our result is novel in the literature, as elaborated in Section \ref{sec:lit}.

\medskip
 
Both Figure  \ref{fig:estK} and our theory indicate an inflation of bias when $p$ becomes excessively large.  We therefore propose  a data-driven way to choose the number of added noise.  If it is believed that most of the informative predictors load strongly on the common factors, then $ p_0\asymp\psi_{p,n}$ and our theory shows:
$$
 \mathbb E[(x_{\new}'\beta-\widehat y_{\new}^*)^2|X]= O_P\left(\frac{p}{p_0n}+\frac{np_0}{p^2}\right).
$$
 We can choose the optimal number of predictors $p$ by minimizing the rate of convergence: 
$$
p =     C\times  (np_0)^{2/3} \asymp \arg\min_p\left(\frac{p}{p_0n}+\frac{np_0}{p^2}\right),
$$
where $C$ is a constant to be determined. 
A data-driven choice of $C$ is available via commonly used tuning strategies, which we now discuss.

For cross-sectional predictions where the ordering of data is unimportant, the ``leave-one-out" is a computationally attractive method. Suppose $C$ is a candidate choice, then denote the total number of predictors, $p(C)=C\times  (np_0)^{2/3}$ (after adding noise).  For each $t\leq n$,  let $\widehat\beta_{-t}(C) $ denote the Ridgeless estimator using $p(C)$ predictors and all data except the $t$-th observation.  Then the optimal $C$ can be chosen by minimizing:
$$
\text{leave-one-out: } \quad  C^*=\arg\min_{C}\sum_{t=1}^n\left( y_t- x_t'\widehat\beta_{-t}(C)\right)^2.
$$
Interestingly, we do not have to compute $n$ leave-one-out $\widehat\beta_{-t}(C)$ for this procedure, thanks to an elegant analytic formula given by \cite{shen2023algebraic}. They showed that the leave-one-out procedure  is equivalent to minimizing: 
\begin{equation}\label{eq4.2loo}
    C^*=\arg\min_C \|  \Diag(G(C))^{-1}G(C) y \|^2,\quad G(C)= (X(C)X(C)')^{-1}
\end{equation}
where $X(C)$ denotes the $n\times p(C)$ matrix of   $p(C)$ predictors, and $\Diag(G)$ takes the diagonal elements of $G$. Here $G(C)$ is invertible because $\rank(X(C))=n<p(C)$, whose computation is manageable so long as $n$ is not very large.

For time-series predictions where there is a natural ordering of the data, we recommend the rolling window tuning for time series cross-validation: predetermine $T_0$ and $V_0$ as the sizes of training and validation sets. Let $\widehat \beta_{t}(C)$ denote the estimated coefficient using $T_0$ periods of data: $t-T_0+1,..., t$, with  tuning parameter $C$. We then choose $C^*$ by:
$$
C^*=\arg\min_C \sum_{l=0}^{L-1}(y_{T_0+l}- x_{T_0+l}'\widehat \beta_{T_0+l}(C))^2.
$$

%----------------------------------------------------------------%
% CV FIGURE %
%----------------------------------------------------------------%
\begin{figure}[h!]
\begin{subfigure}[t]{0.48\textwidth}
  \centering
  \includegraphics[width=\textwidth]{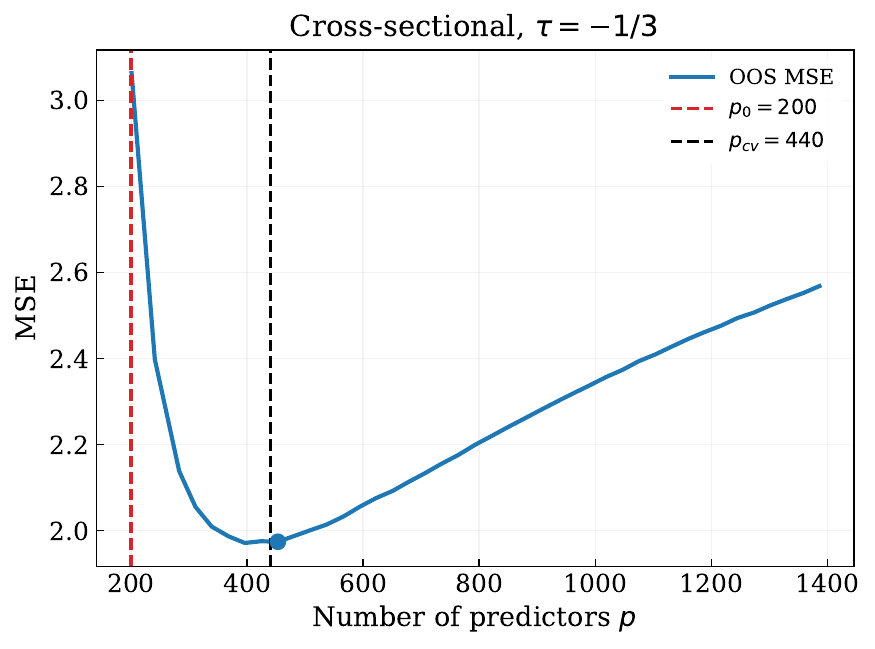}
  \caption{Cross-sectional prediction}
\end{subfigure}\hfill
\begin{subfigure}[t]{0.48\textwidth}
  \centering
  \includegraphics[width=\textwidth]{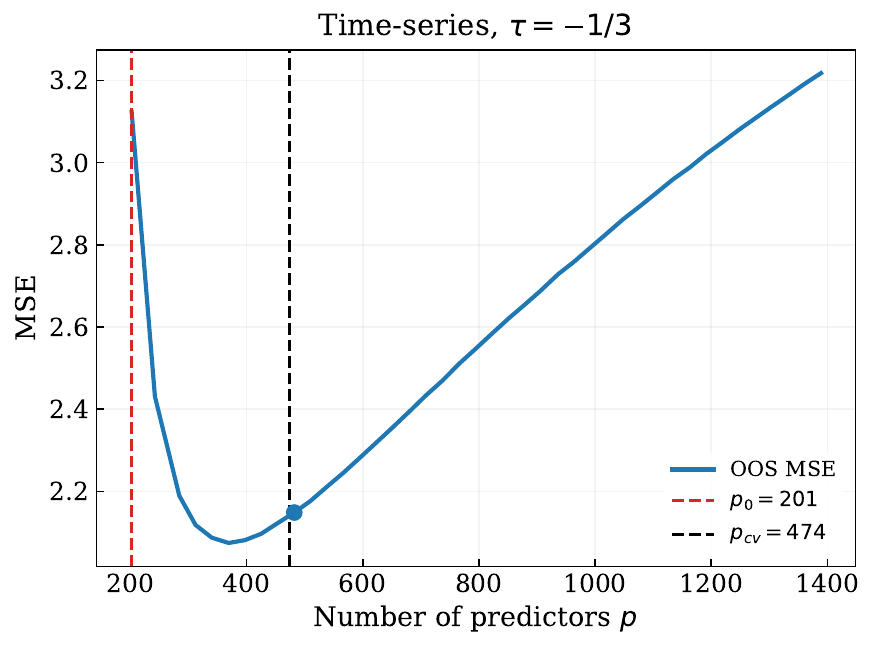}
  \caption{Time-series prediction}
\end{subfigure}
\caption{Predictive MSE and Data-driven Tuning}\label{fig:simcv}

\begin{flushleft}
\footnotesize
Notes: Out-of-sample MSE of the noise-regularized Rao-Blackwell estimator $\widehat y_{\new}^* = x_{\new, I}'\widehat\beta_I^*$ as the total number of predictors $p$ increases. The first $p_0$ predictors are informative and generated from the factor model in Section~\ref{sec:simulation} with signal strength $\tau = -1/3$; the remaining $p-p_0$ are i.i.d.\ $N(0,1)$ noise. The sample size is $n = 100$. Panel~(a) is a cross-sectional setting where factors are i.i.d.\ sequences, and $p_{cv}$ is selected by Rao-Blackwell leave-one-out cross-validation \eqref{eq4.2loo}. Panel~(b) is a time-series setting where factors follow a stationary AR(1) process, and $p_{cv}$ is selected by rolling-window cross-validation with the first half of the sample reserved for tuning. The dashed vertical line marks $p_{cv}$ averaged over 50 replications.
\end{flushleft}
\end{figure}

To illustrate the effectiveness of data-driven choices of $p_{cv}=p(C^*)$, Figure \ref{fig:simcv} plots the predictive MSE of two simulated designs, where the DGP is generated as in one of the simulation designs in Section \ref{sec:sim_unknown}  with weak factors.  The first $p_0=200$ predictors are informative predictors, whereas the rest $p-p_0$ predictors are intentionally added noise variables. In both cases the data-driven choice $p_{cv}$ suitably identifies a proper number of added noise.

\subsection{Relation with the linear double descent literature}
\label{sec:lit}
Contrary to conventional statistical wisdom, which asserts that overfitting undermines forecast performance by inflating out-of-sample variance,  the double descent phenomenon shows that it is possible to retain predictability by accumulating the number of regressors beyond the threshold of interpolation. 

It has gained increasing attention in the machine learning community. As the model complexity exceeds the sample size and continues to grow,  a second descent of the prediction occurs in the excessively overparametrized regime. It  was first  illustrated in the empirical work by \citet{belkin2019reconciling}, \citet{hastie2022surprises}, and  \citet{arora2019implicit}, and its theory has been explored in linear models with Ridgeless regressions, e.g., \citet{mei2019generalization, belkin2020two, dobriban2018high, lee2023mean}.

Our work differs from the recent statistical literature on overparameterized models in three important respects. First, while the literature does not explicitly assume $\|\beta\|\to 0$, this condition is in fact necessary for the forecast consistency.
In the i.i.d.~setting without presence of strong factors,  \cite{hastie2022surprises} characterize the limiting predictive MSE, whose convergence requires  this condition.  It turns out that  $\|\beta\|\to 0$  remains   necessary in the presence of factors; otherwise, the overparametrized bias would not vanish asymptotically.  
However, the literature has   provided little justification for why such a condition should hold in a practical model.

Our paper provides a formal and economically motivated justification for this condition. Our true DGP is driven by a low-dimensional factor structure  $y_t = \rho' f_t + \epsilon_{y,t},$ 
while the econometrician estimates an overparameterized linear regression  $
y_t = x_t' \beta + u_t
$ 
as a working model. 
In this environment, the population regression coefficient $\beta$ is a reduced-form object implied by the underlying factor structure. 
As the cross-sectional dimension grows, the factor structure  implies $\|\beta\|\to 0$. Hence, the shrinkage condition required for consistency is not imposed, but arises  naturally from a standard economic DGP. 

Secondly, as explained in the previous subsection, we show that the second descent arises from the shrinkage of the eigenvalue spectrum induced by the presence of many noise variables. Although these noise variables contain no predictive information about the target variable, they cause the high-dimensional sample covariance matrix to have a highly dispersed eigenvalue spectrum. This lowers the out-of-sample variance.

Lastly, from an important technical perspective, our framework explicitly analyzes the impact of a few very large eigenvalues on overparametrized forecasts.  In much of the recent work, the design matrix is assumed to have uniformly bounded eigenvalues, which effectively imposes a weak dependence structure across predictors.  In contrast, many economic forecasting environments are inherently driven by a small number of latent common factors. These factors generate strong cross-sectional dependence among informative predictors, inducing an approximate low-rank structure in the population covariance matrix. While some recent work on overparameterized models allows for diverging eigenvalues induced by factor structures (e.g., \cite{meng2025estimation}), the implications of such growing eigenvalues in forecasting  models remain unclear.

 We provide a careful analysis of this case and show how the interaction between factor-driven spectral spikes and overparameterization shapes the predictive risk. To understand the intuition, start with the matrix form of the factor model
\begin{equation}\label{facmatrix}
\underset{(n\times p)}{X}= \underset{(n\times K)}{F} \underset{(K\times p)}{\Lambda'}  + U.
\end{equation}
By applying   SVD on both $X'X$ and $\mathbb E x_tx_t'$, we can separate the effect of the factor model   from its remaining counterparts. First, consider the SVD of $\mathbb Ex_tx_t' $: 
$$\mathbb Ex_tx_t' = \bar V_{K} \bar D_K^2\bar V_{K}'+\bar V_{-K} \bar D_{-K}^2\bar V_{-K}' $$  
corresponding to the first $K$ eigenvectors/values  $(\bar V_{K}, \bar D_K)$ and  last $p-K$ eigenvectors/values  $(\bar V_{-K}, \bar D_{-K})$. In addition, consider the SVD on $X'X$: 
$$
X'X = V_KD_K^2 V_K' + V_{-K}D_{-K}^2 V_{-K}' + V_{-n} 0 V_{-n}',
$$
where $D_K^2$ and $D^2_{-K}$ respectively denote the diagonal matrices of the largest $K$ and the remaining $n-K$ nonzero eigenvalues of $X'X$; columns of $V_K$ and $V_{-K}$   are the corresponding eigenvectors; columns of $V_{-n}$  are the eigenvectors corresponding to the  $p-n$ zero-eigenvalues.

Then we  decompose the forecast bias and variance of $\widehat y_{\new}$ into:
\begin{eqnarray*}
 \mathbb E [(x_{\new}' \delta_\beta)^2 |X ]&=& A_1 + A_2,\quad \mathbb E [ (x_{\new}'B_{X}\epsilon_{y})^2 |X_I] = B_1+ B_2 
\end{eqnarray*}
where for $\Sigma_X = \mathbb E (x_t x_t')$, 
\begin{eqnarray*}
    A_1=\underbrace{\| \bar V_{K} \bar D_K\bar V_{K}' V_{-n}V_{-n}'\beta\|^2}_{\text{factor-model design}}
, && B_1= \underbrace{\sigma_e^2\tr\left(V_KD_K^{-2}V_{K}'\Sigma_X\right)}_{\text{factor-model design}}, \\
A_2=\underbrace{\| \bar V_{-K} \bar D_{-K}\bar V_{-K}' V_{-n}V_{-n}'\beta\|^2}_{\text{i.i.d.~design}}, 
  && 
B_2= \underbrace{\sigma_e^2\tr\left(V_{-K}D_{-K}^{-2}V_{-K}'\Sigma_X\right)}_{\text{i.i.d.~design}}
\end{eqnarray*}
Both bias and variance depend on components that arise from the factor model $A_1, B_1$, and from the idiosyncratic noise $A_2, B_2$ that behave similarly to the classic i.i.d.~design. In particular, the convergence of $A_1$ and $B_1$ rely on three properties of the factor model: (i) the population eigenvectors $\bar V_K$ can be well approximated by the sample eigenvectors $V_K$; (ii) the implied reduced form parameter is dense in that $\|\beta\|\to0 $; and (iii) the top $K$ eigenvalues grow very fast. As for $A_1$, the sample eigenvectors corresponding to the first $K$ eigenvalues are orthogonal to $V_{-n}$. Hence  (i) implies $\|  \bar V_{K}'V_{-n}\|\approx \| V_{K}'V_{-n}\|  \stackrel{P}{\to} 0$,
and the convergence of $A_1$ is ensured together with (ii). In addition, the behavior of $B_1$ is driven by the top $K$ sample eigenvalues, and thus $B_1$ converges as $\|D_K^{-2}\|= O_P((\psi_{p,n}n)^{-1})$.

In summary, when most of the high dimensional predictors are generated by a few common factors,  the factors lead to  a few spiked eigenvalues facilitating the convergence of variance; they also naturally yield a dense regression coefficient to shrink the bias.

%---------------------------------------------------------------------------------------%
% SIMULATION 
%---------------------------------------------------------------------------------------%

\section{Simulation} \label{sec:simulation}

We run Monte Carlo experiments to check the performance of noise-regularization under the factor model specification in \eqref{eq2.1} and \eqref{eq2.2}. The outcome variable is generated from a $K=4$ factor model $y_t = \rho' f_t + \epsilon_{y,t}$, with $p_0$ informative predictors and $p - p_0$ noise:
$$
x_{i,t} = \begin{cases}
\lambda_i' f_t + u_{i,t}, & i = 1, \ldots, p_0, \\
u_{i,t}, & i = p_0 + 1, \ldots, p,
\end{cases}
\quad \text{where} \quad \lambda_i = \lambda_{i,0} \times p_0^{\tau},
$$
and $(f_t, \epsilon_{y,t}, \lambda_{i,0}, u_{i,t})$ are all standard normal, and $\rho \sim N(0, I_K)$. Here $\tau \in [-1/2, 0]$ determines the factor strength so that $\Lambda_I' \Lambda_I \asymp p_0^{1+2\tau}$; the closer $\tau$ to zero, the stronger the factors. To capture heterogeneous factor strengths, we allow $\tau$ to differ across the $K = 4$ factors and consider two configurations:
\begin{itemize}[nosep]
\item \emph{Homogeneous:} $\tau = -1/3$ for all 4 factors (moderately weak).
\item \emph{Mixed:} $\tau = (-1/4,\, -1/3,\, -1/2,\, -1/2)$, combining one strong, one moderate, and two weak factors.
\end{itemize}
The mixed case is of particular interest because the two factors with $\tau = -1/2$ are very weak. The sample size is $n = 100$ throughout, and test MSE is evaluated on $n_{\new} = 500$ new observations. All results are averaged over 100 Monte Carlo replications. We set $p$ to take values on a grid evenly spaced from 1 to $p_{\max} = 1000$.
We consider the two scenarios as in Sections \ref{sec:unknown} and \ref{sec:known}, respectively.

\subsection{Scenario I: unknown noise identities} \label{sec:sim_unknown}

\begin{figure}[h!]
\centering
\begin{subfigure}[t]{0.48\textwidth}
  \centering
  \includegraphics[width=\textwidth]{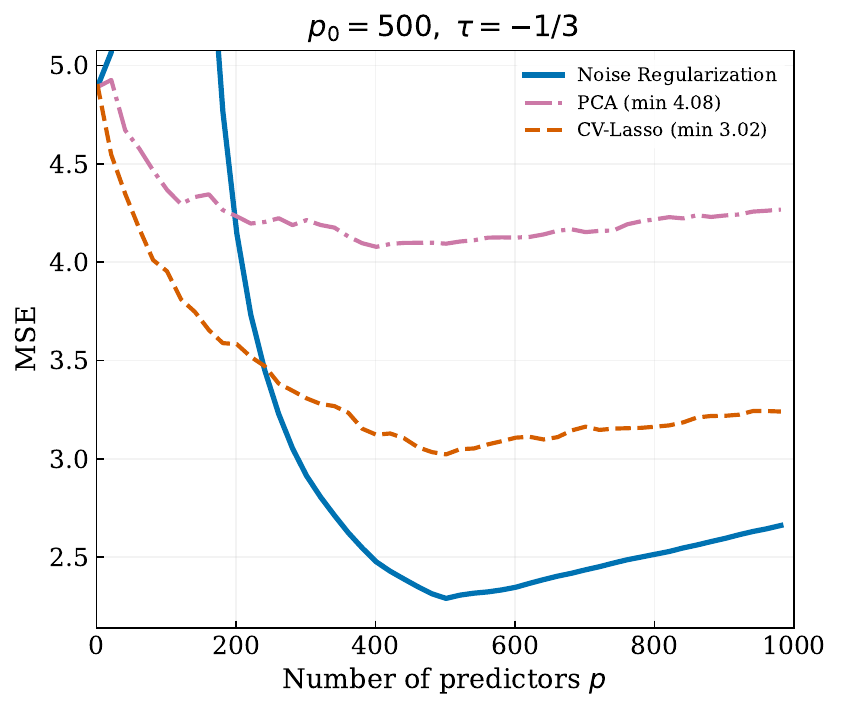}
  \caption{$p_0 = 500$, $\tau = -1/3$}
\end{subfigure}\hfill
\begin{subfigure}[t]{0.48\textwidth}
  \centering
  \includegraphics[width=\textwidth]{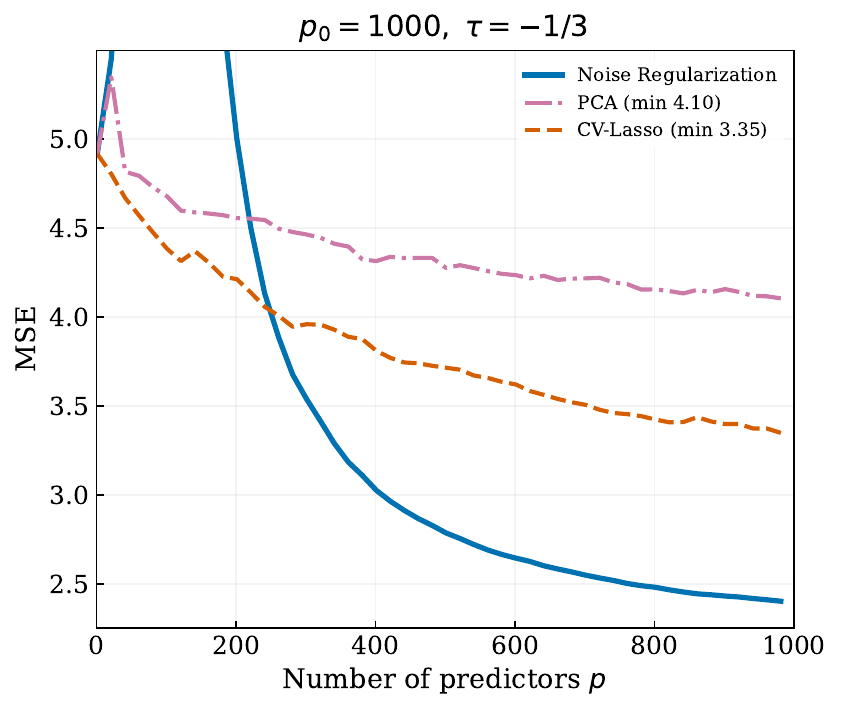}
  \caption{$p_0 = 1000$, $\tau = -1/3$}
\end{subfigure}

\vspace{0.5em}
\begin{subfigure}[t]{0.48\textwidth}
  \centering
  \includegraphics[width=\textwidth]{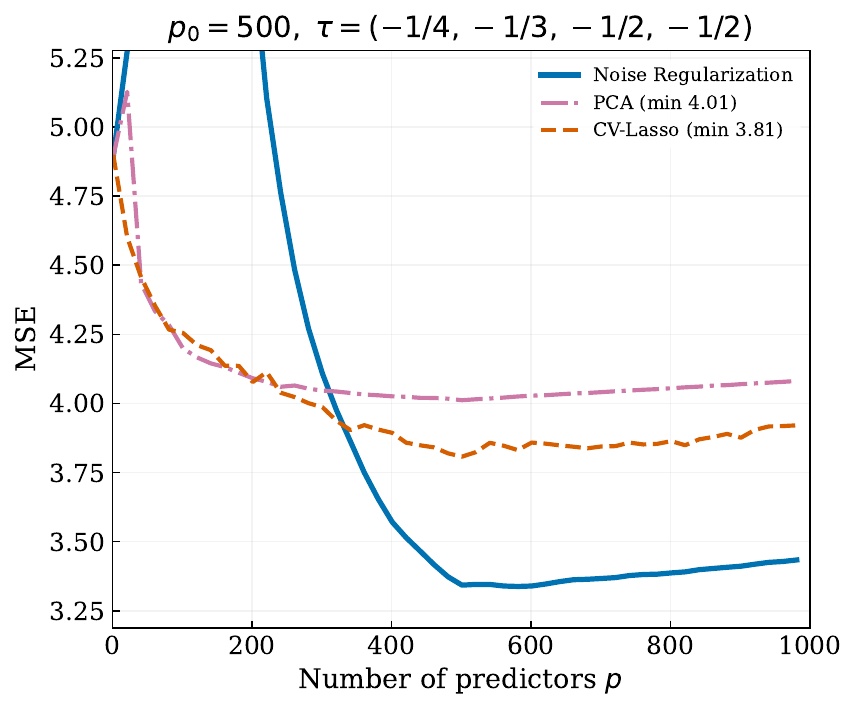}
  \caption{$p_0 = 500$, mixed $\tau$}
\end{subfigure}\hfill
\begin{subfigure}[t]{0.48\textwidth}
  \centering
  \includegraphics[width=\textwidth]{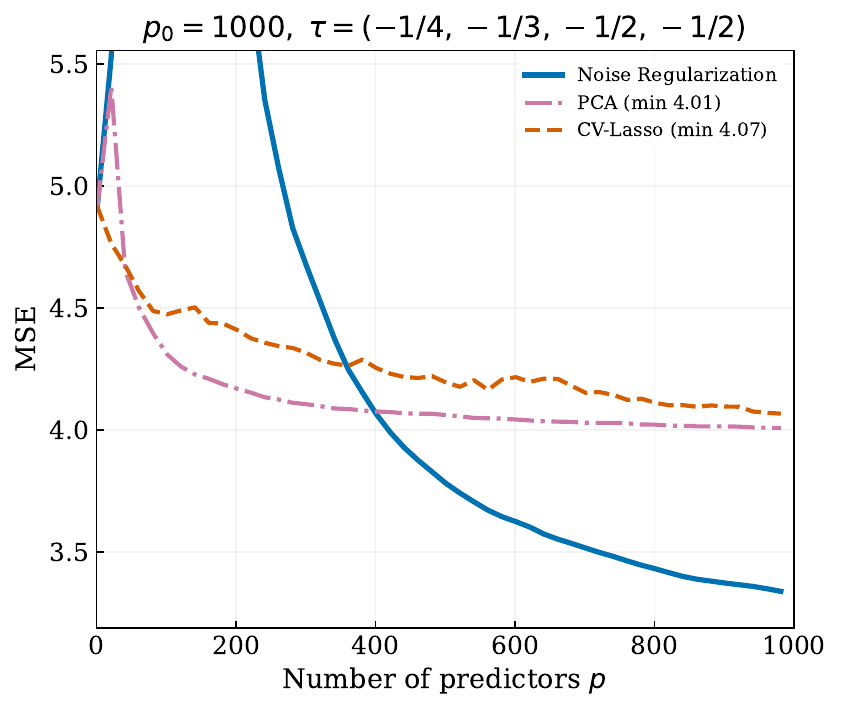}
  \caption{$p_0 = 1000$, mixed $\tau$}
\end{subfigure}
\caption{Scenario I: predictive MSE as the number of predictors $p$ increases. The sample size is $n = 100$. All three methods use the full set of $p$ predictors. The top row has homogeneous factor strength $\tau = -1/3$; the bottom row mixes $\tau = (-1/4, -1/3, -1/2, -1/2)$. PCA and CV-Lasso minimum MSE values are reported in the legend when their curves exceed the plotted range.}
\label{fig:sim1}
\end{figure}

The economist uses all $p$ collected predictors without knowing which predictors are informative. We set $p_0 \in \{500, 1000\}$; all predictors are informative when $p \leq p_0$, and the additional $p - p_0$ are noise when $p > p_0$. Three methods are compared, all using the full set of $p$ predictors: the proposed noise-regularized estimator, PCA with the number of factors chosen by  \cite{BN02}, and CV-Lasso where the tuning parameter is chosen by 10-fold cross-validation.  

Figure \ref{fig:sim1} plots the predictive MSE as $p$ increases.
All predictors are informative when $p \leq p_0$, whereas $p - p_0$ noise variables enter when $p > p_0$. In all four panels, the noise-regularization dominates both PCA and CV-Lasso, particularly in the overparameterized regime ($p > n$).
In panel (b), where all predictors are informative ($p_0 = 1000$) with a homogeneous $\tau = -1/3$, including more weak predictors continuously benefits noise-regularization, driving its MSE down to 2.40 at the right edge. In panels (a) and (c), where $p_0 = 500$ and the number of informative predictors stops increasing, the MSE curve flattens after $p_0$ but remains well below the competitors.

PCA gains little from a large $p$ because the factors are relatively weak, which misses a substantial portion of the predictive signal. CV-Lasso struggles for a different reason: the dense factor structure spreads the signal across all informative predictors, each carrying a small but nonzero coefficient. Lasso's variable selection, designed for sparse models, selects too few predictors to recover this diffuse signal.

\subsection{Scenario II: intentionally added noise} \label{sec:sim_known}

When the economist knows which predictors are informative, she intentionally adds noise to implement noise-regularization. The $p_0$ informative predictors are augmented with $p - p_0$ columns of $N(0,1)$ noise. The  noise-regularized estimator is computed  for  $B = 50$ times with independent noise draws, and the predictions are averaged.

We set $p_0 \in \{200, 300\}$ and compare four methods: Noise-regularized with the number of added noise variables  selected via leave-one-out cross-validation;\footnote{Algorithm \ref{app:rbloo} in Section \ref{sec:implementation} of the appendix presents a detailed algorithm for selecting the number of added noise using leave-one-out method.}  CV-Ridge which applies the Ridge regression on the $p_0$ informative predictors and tuned by 10-fold cross-validation; CV-Lasso and PCA.
Except for noise-regularization, all the three competing methods use the informative predictors only: they are oracle methods with  perfect variable selection.

\begin{figure}[h!]
\centering
\begin{subfigure}[t]{0.48\textwidth}
  \centering
  \includegraphics[width=\textwidth]{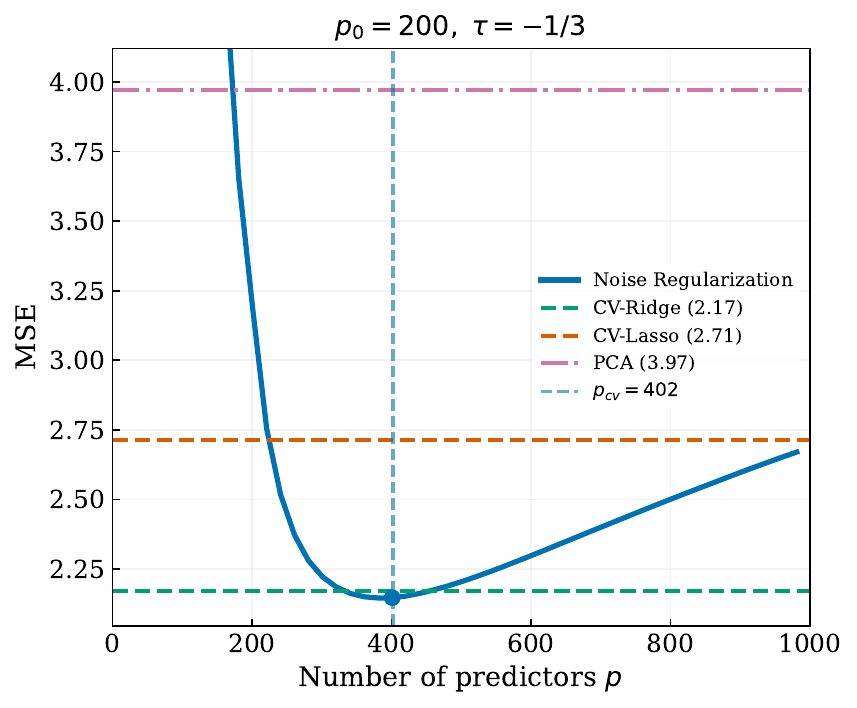}
  \caption{$p_0 = 200$, $\tau = -1/3$}
\end{subfigure}\hfill
\begin{subfigure}[t]{0.48\textwidth}
  \centering
  \includegraphics[width=\textwidth]{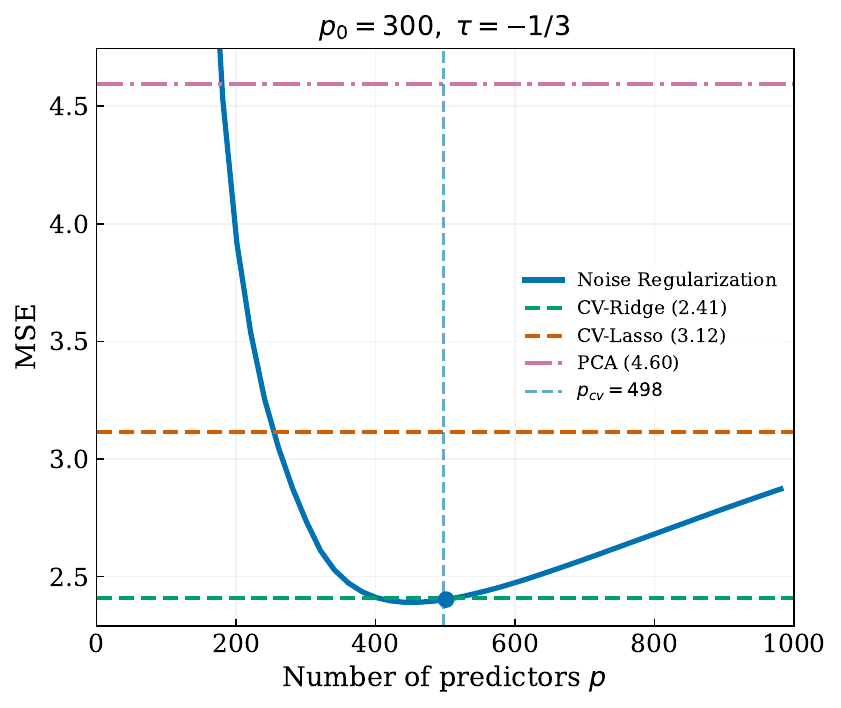}
  \caption{$p_0 = 300$, $\tau = -1/3$}
\end{subfigure}

\vspace{0.5em}
\begin{subfigure}[t]{0.48\textwidth}
  \centering
  \includegraphics[width=\textwidth]{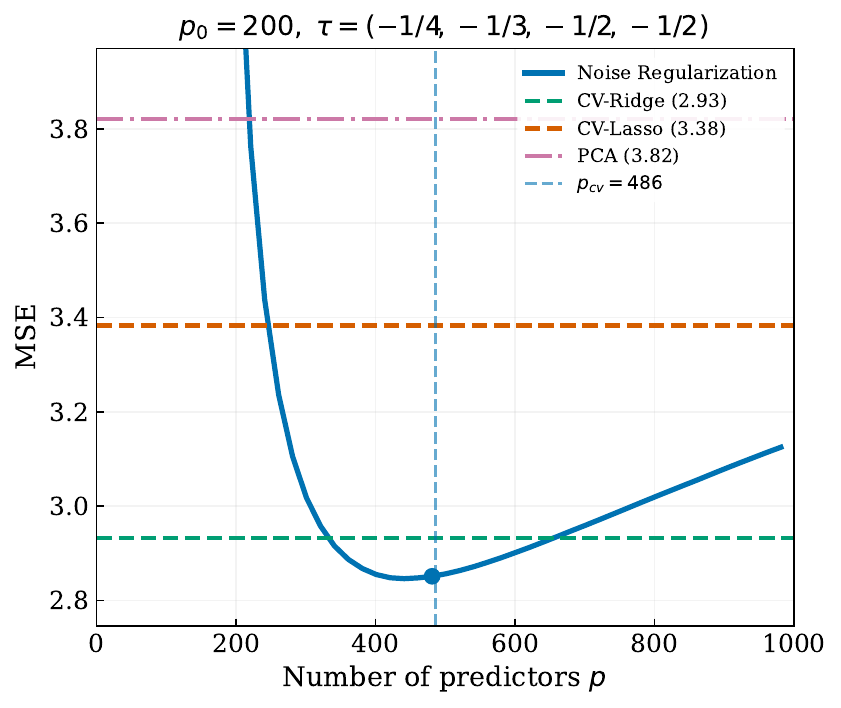}
  \caption{$p_0 = 200$, mixed $\tau$}
\end{subfigure}\hfill
\begin{subfigure}[t]{0.48\textwidth}
  \centering
  \includegraphics[width=\textwidth]{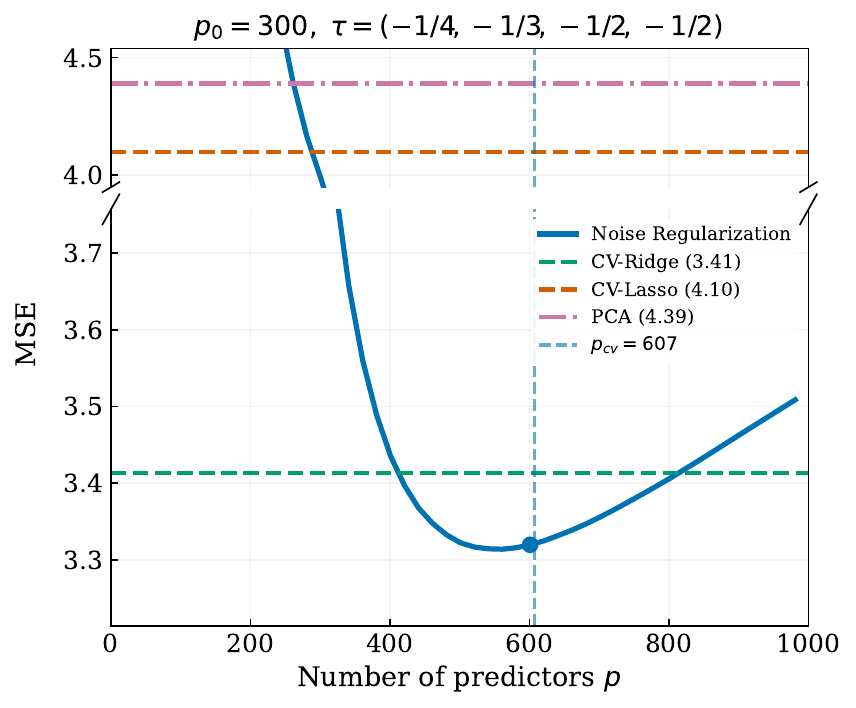}
  \caption{$p_0 = 300$, mixed $\tau$}
\end{subfigure}
\caption{Scenario II: predictive MSE as the number of predictors $p$ increases. The sample size is $n = 100$; the noise-regularized estimator averages $B = 50$ Rao-Blackwell draws. The vertical dashed line marks $p_{cv}$ selected by leave-one-out cross-validation. CV-Ridge, CV-Lasso, and PCA use only the $p_0$ informative predictors (oracle variable selection).}
\label{fig:sim2}
\end{figure}

Figure \ref{fig:sim2} plots the predictive MSE as $p$ increases. The horizontal dashed lines show the MSE of the oracle methods, and the solid curve is the noise-regularized forecast. The vertical dashed line indicates $p_{cv}$ selected by leave-one-out cross-validation. The  cross-validation selects $p_{cv}$ close to the oracle optimum. In panel (a), for instance, we select  $p_{cv} \approx 400$ while the minimum of the MSE curve occurs at $p \approx 380$. 

In all four panels, noise-regularization at $p_{cv}$ outperforms CV-Ridge, despite the latter's access to the oracle. The improvement is most pronounced in the mixed-$\tau$ panels: with $p_0 = 200$, the noise-regularized MSE is 2.85 versus 2.93 for CV-Ridge; with $p_0 = 300$, the gap widens to 3.32 versus 3.41. Even in the homogeneous case ($\tau = -1/3$), noise-regularization slightly outperforms Ridge (2.15 vs.\ 2.17 for $p_0 = 200$; 2.40 vs.\ 2.41 for $p_0 = 300$).

The advantage over Ridge is amplified when factor strengths are heterogeneous. Ridge regression applies a single penalty $\lambda$ to all coefficients, whereas the implicit regularization through noise addition adapts to the signal structure---coefficients associated with stronger factors are shrunk less. This heterogeneous shrinkage is a consequence of the randomized regularization mechanism and cannot be replicated by a single Ridge penalty. CV-Lasso and PCA both perform substantially worse. Lasso suffers from the dense factor structure, while PCA fails to detect the weakest factors, especially in the mixed-$\tau$ panels where two factors have $\tau = -1/2$.

%------------------------------------------------------------------%
% NEW EMPIRICS
%------------------------------------------------------------------%

\section{Empirical applications} \label{sec:empirics}

We apply the noise-regularized forecast to three settings in which high-dimensional prediction is of first-order importance: U.S.\ inflation, cross-country GDP growth, and the U.S.\ equity premium. In each application the economist  observes $p_0$ economic predictors and augments them with $p - p_0$ columns of i.i.d.~$N(0,1)$ noise. The Rao-Blackwell denoised prediction averages $B = 50$ independent noise draws. The total dimension $p$ is selected by data-driven cross-validation, specifically leave-one-out in the cross-sectional application (growth) and rolling-window time-series CV in the two time-series applications (inflation and equity premium). We compare with CV-Lasso and CV-Ridge, both tuned by five-fold cross-validation and using only the $p_0$ economic predictors. 

In the subsequent subsections, we study the three applications in detail. Table~\ref{tab:summary} summarizes the main results across the studies.

\begin{table}[H]
\centering
\begin{threeparttable}
\caption{Summary of empirical results}
\label{tab:summary}
\small
\begin{tabular}{lrrrrrr}
\toprule
Application & $p_0$ & $p_{cv}$ & Noise-reg & CV-Lasso & CV-Ridge & PCA \\
\midrule
\multicolumn{7}{l}{\emph{Predictive MSE }} \\[3pt]
U.S.\ inflation (Section~\ref{sec:inflation})  & 102 & 526   & \textbf{0.982} & 1.103 & 1.102 & 1.122 \\
Growth (Section~\ref{sec:growth})               & 60  & 396   & \textbf{0.940} & 1.262 & 1.193 & 1.073 \\[6pt]
\multicolumn{7}{l}{\emph{Out-of-sample $R^2$ (\%) (Section~\ref{sec:equity_premium})}} \\[3pt]
Equity premium (annual)    & 15 & 1{,}000  & $\mathbf{+1.81}$ & $-14.49$ & $-11.92$ & $-173$\tnote{a} \\
Equity premium (quarterly) & 15 & 4{,}000  & $\mathbf{+1.08}$ & $-4.55$  & $-2.29$  & $-35.5$ \\
Equity premium (monthly)   & 15 & 12{,}000 & $\mathbf{+0.37}$ & $-0.46$  & $-1.59$  & $-8.8$ \\
\bottomrule
\end{tabular}
\begin{tablenotes}                                             \small 
\item[a] PCA performance for the annual equity premium prediction is particularly poor due to the small sample size ($n=20$) relative to the number of predictors ($p_0=15$). Even with the maximum number of factors capped at $K_{\max}=10$, the Bai--Ng information criterion selects the maximum, leaving few degrees of freedom and producing unstable out-of-sample forecasts.                                           
\end{tablenotes}
\end{threeparttable}
\end{table}

\subsection{U.S.\ inflation forecast} \label{sec:inflation}

Accurate inflation forecasts are essential for monetary policy, business planning, and investment decisions. We use the FRED-MD dataset to forecast the change in the U.S.\ inflation rate 
$$
\Delta \text{inflation}_{t+1}= \log(\text{CPI}_{t+1} /\text{CPI}_{t}  ) -\log(\text{CPI}_{t} /\text{CPI}_{t-1} ),
$$
as defined by \citet[p.580]{mccracken2016fred}. 
The dataset includes macroeconomic predictors spanning June 1959 to January 2024.  We exclude variables with  missing observations over this period, resulting in  $p_0 = 102$ predictors. We use 120-month (10-year) rolling windows to produce one-month-ahead forecasts. This dataset is widely recognized for its inherent challenge of relatively weak factors: data-driven criteria such as \cite{BN02}'s information criteria  typically suggests 8--10 factors, explaining only 50--62\% of total variation.

Adding noise predictors inflates the eigenvalues of the design matrix and provides the implicit regularization that avoids estimating  weak factors. 
Noise-regularization augments the $p_0$ macro variables with $p - p_0$ columns of $N(0,1)$ noise, and the dimension $p$ is selected by time-series cross-validation on a burn-in period prior to the out-of-sample evaluation. The optimal number of predictors is $p_{cv} = 526$, corresponding to roughly 420 added noise columns. CV-Lasso, CV-Ridge, and PCA use only the original macro predictors throughout.

\begin{figure}[h!]
\centering
\includegraphics[width=0.65\textwidth]{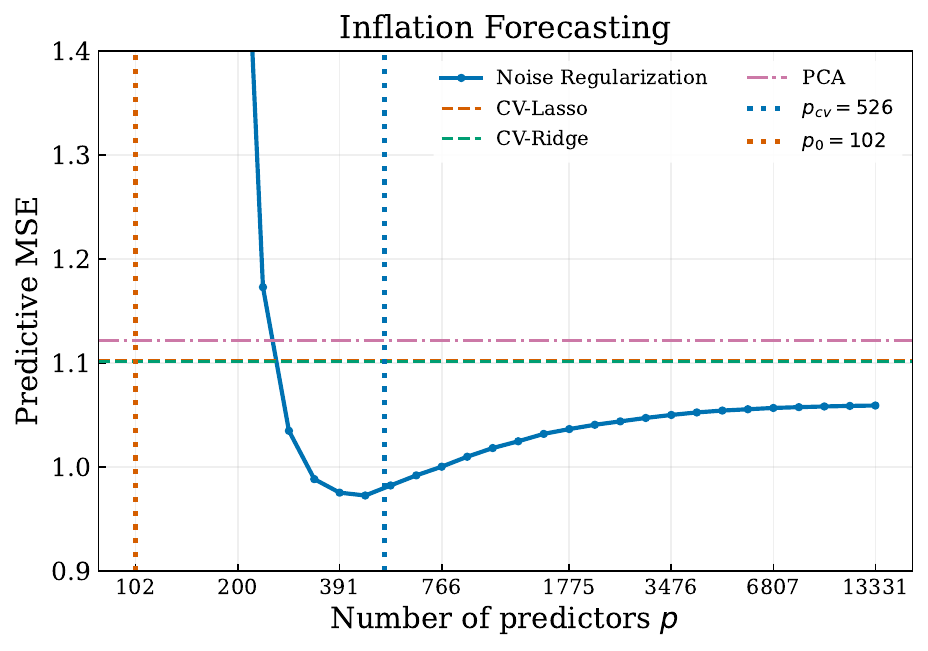}
\caption{Inflation forecasting: predictive MSE as the number of predictors $p$ increases. The $p_0 = 102$ FRED-MD macro predictors are augmented with $p - p_0$ noise columns. Rolling 120-month windows, one-month-ahead forecast. The vertical dashed line marks $p_{cv} = 526$ selected by time-series CV. CV-Lasso and CV-Ridge use only the original macro predictors.}
\label{fig:inflation}
\end{figure}

Figure~\ref{fig:inflation} plots the predictive MSE as $p$ increases on a logarithmic grid. The noise-regularized forecast exhibits a sharp spike near the interpolation threshold ($p \approx n = 120$) but its MSE drops rapidly as more noise is introduced, outperforming benchmarks by $p \approx 400$. At $p_{cv} = 526$ the noise-regularized MSE is 0.982, an 11\% improvement over CV-Lasso (1.103) and CV-Ridge (1.102). Beyond $p_{cv}$, the MSE curve remains nearly flat through $p > 13{,}000$, demonstrating the robustness to the number of added noise variables predicted by the theory.

\subsection{Growth forecasts} \label{sec:growth}
Economic growth forecasting has been one of the earliest high-dimensional applications in economics \citep{sala1997just}. We use the dataset of \cite{barro1994sources}, which contains $p_0 = 60$ socio-economic and geographical characteristics for 90 countries. In each replication we randomly split the countries into a training set of $n = 45$ and a test set of 45, fit the model on the training set, and evaluate predictions on the test set. Results are averaged over 50 random splits. 
The noise-regularized method augments the 60 predictors with $p - p_0$ noise columns. The selected total number of predictors, including noise, is   $p_{cv} = 396$ using leave-one-out cross-validation.

\begin{figure}[h!]
\centering
\includegraphics[width=0.65\textwidth]{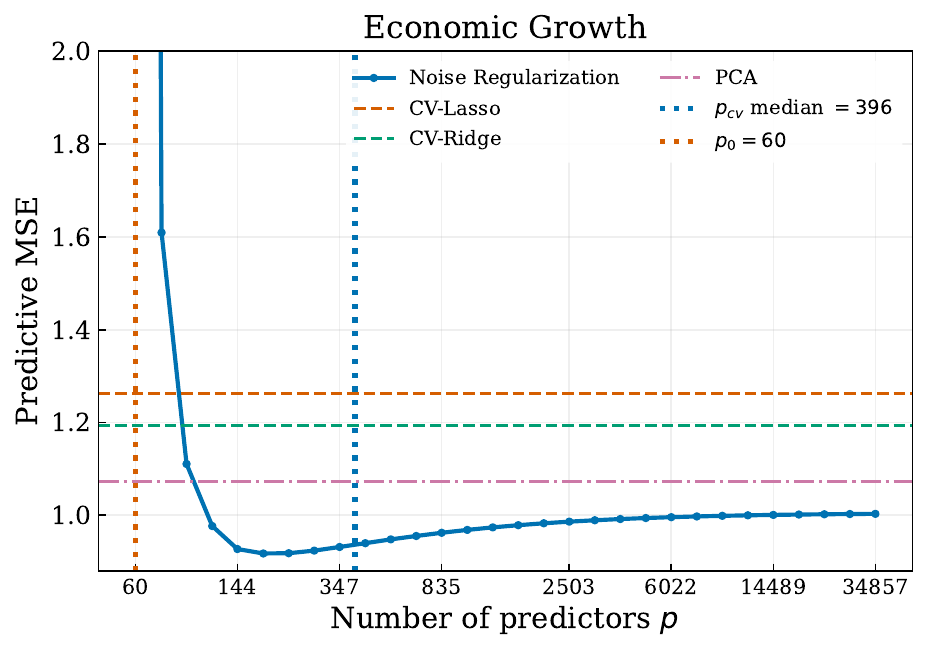}
\caption{Growth forecasting: predictive MSE as the number of predictors $p$ increases. The $p_0 = 60$ socio-economic variables from \cite{barro1994sources} are augmented with $p - p_0$ noise columns. Random 50/50 training/test splits, averaged over 50 replications. The vertical dashed line marks $p_{cv} = 396$. CV-Lasso and CV-Ridge use only the original 60 predictors. PCA uses \cite{BN02} factor selection with $K_{\max}=10$.}
\label{fig:growth}
\end{figure}

Figure~\ref{fig:growth} plots the predictive MSE. The MSE using only the original 60 predictors via Ridgeless is very large ($\approx 14$) due to overfitting in the $p \approx n$ regime. After the interpolation threshold, the MSE drops precipitously as noise dimensions are added and stabilizes below 1.0 by $p \approx 200$. At $p_{cv} = 396$ the noise-regularized MSE is 0.940, a 25\% improvement over CV-Lasso (1.262) and a 21\% improvement over CV-Ridge (1.193). PCA achieves an MSE of 1.073, outperforming Lasso and Ridge but still 14\% above noise regularization. The MSE remains stable through $p > 34{,}000$, i.e., nearly 600 times the original predictor count, underscoring the method's robustness.

\subsection{U.S.\ equity premium prediction} \label{sec:equity_premium}

Predictability of the U.S.\ equity premium is a central and contentious question in asset pricing. \cite{Goyal2008comprehensive} conducted a comprehensive out-of-sample examination of prevailing predictive models, concluding that most fail to beat the historical-mean benchmark. \cite{goyal2023comprehensive} updated the analysis with new predictors and methods, arriving at qualitatively similar conclusions: the majority of models yield negative or barely positive out-of-sample $R^2$.

We revisit this question using the 15 predictors in \cite{Goyal2008comprehensive}: dividend-price ratio, earnings-price ratio, book-to-market, Treasury bill rate, default yield spread (AAA, BAA), long-term yield, net equity expansion, risk-free rate, inflation, long-term return, corporate bond return, stock variance, and lagged market returns. We forecast the equity premium at annual, quarterly, and monthly horizons.\footnote{We drop the cross-sectional premium (csp) because it is discontinued after 2002; the remaining 15 predictors extend through 2024.} The out-of-sample evaluation covers 1960--2024 using 20-year rolling windows, harmonized across all three frequencies. The out-of-sample $R^2$ is defined as
$$
R^2_{\text{oos}} = 1 - \frac{\sum_{t}(y_{t+1}-\widehat y_{t+1})^2}{\sum_{t}(y_{t+1}-\bar y_t)^2},
$$
where $\bar y_t$ is the (expanding window) historical mean of returns up to time $t$.

At each out-of-sample date, we augment the 15 predictors with $p - 15$ columns of $N(0,1)$ noise, compute the noise-regularization method averaged over $B=50$  independent draws as Rao-Blackwellization. The dimension $p$ is selected by time-series cross-validation on a burn-in period (pre-1960 data).

\begin{figure}[h!]
\centering
\begin{subfigure}[t]{0.48\textwidth}
  \centering
  \includegraphics[width=\textwidth]{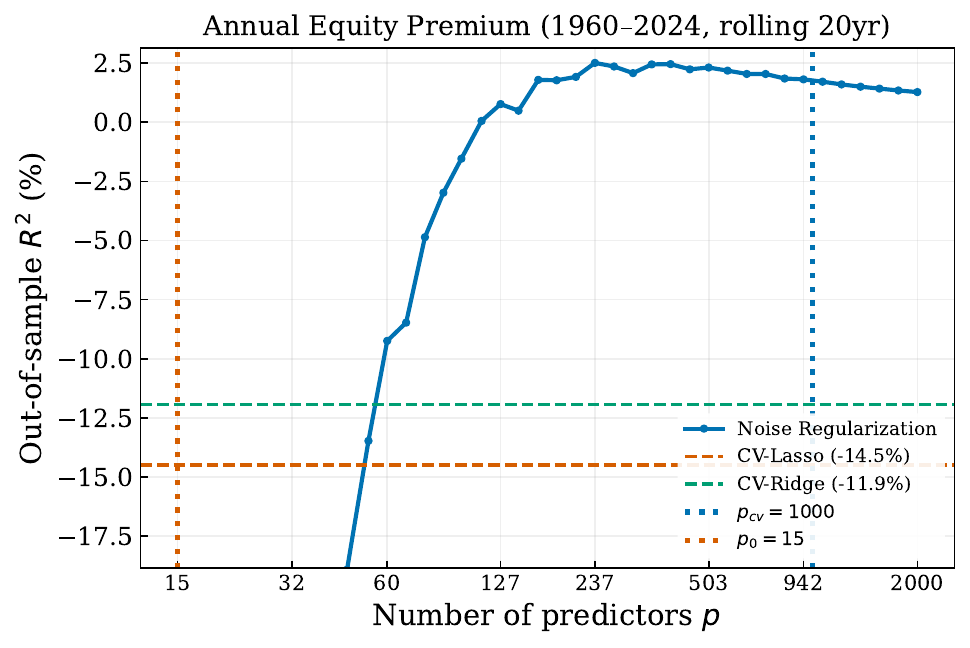}
  \caption{Annual}
  \label{fig:finance_annual}
\end{subfigure}\hfill
\begin{subfigure}[t]{0.48\textwidth}
  \centering
  \includegraphics[width=\textwidth]{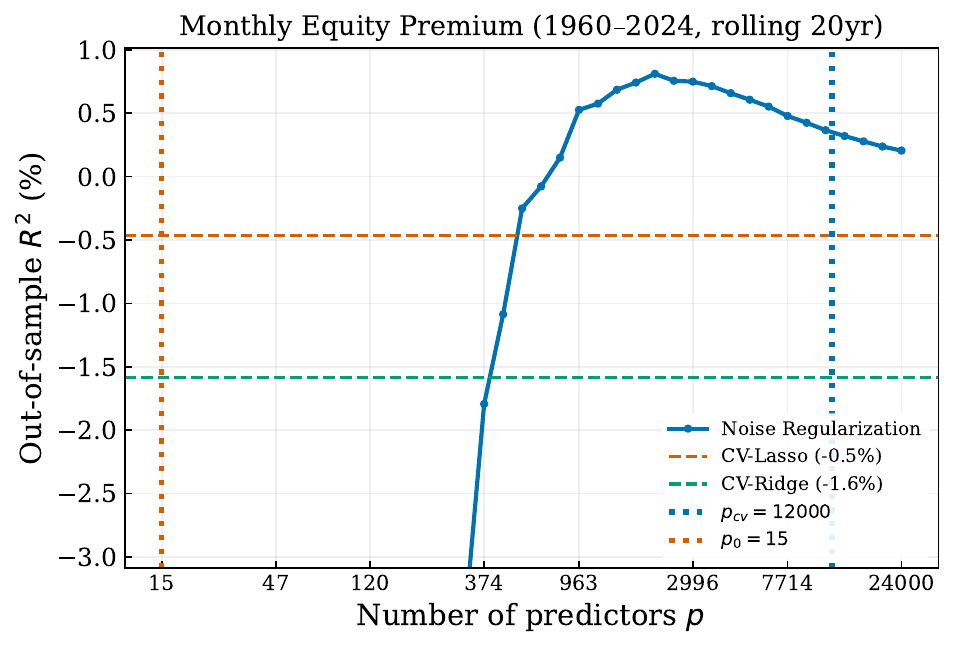}
  \caption{Monthly}
  \label{fig:finance_monthly}
\end{subfigure}
\caption{Out-of-sample $R^2$ (\%) for predicting the U.S.\ equity premium at annual and monthly horizons, 1960--2024, using 20-year rolling windows. The solid curve shows the noise-regularized forecast as $p$ increases (15 economic predictors plus $p - 15$ noise columns, averaged over $L = 50$ Rao-Blackwell draws). Horizontal dashed lines show CV-Lasso (red) and CV-Ridge (green), both using only the 15 economic predictors. The vertical dotted lines mark $p_0 = 15$ and $p_{cv}$ selected by time-series CV. PCA with \cite{BN02} factor selection ($K_{\max}=10$) is omitted as its $R^2$ falls far outside the displayed range (see Table~\ref{tab:summary}). Quarterly results are in Appendix~\ref{app:quarterly}.}
\label{fig:finance}
\end{figure}

Figure~\ref{fig:finance} displays the results. For small $p$, the noise-regularized forecast has strongly negative $R^2$ because the augmented design is near or below the interpolation threshold ($p \approx n$). As $p$ grows into the overparameterized regime, the $R^2$ rises sharply and stabilizes in the positive territory: precisely the second-descent behavior predicted by the theory.

At the annual frequency (panel~a), the noise-regularized forecast achieves $R^2_{\text{oos}} = +1.81\%$ at $p_{cv} = 1{,}000$, compared with $-14.49\%$ for CV-Lasso and $-11.92\%$ for CV-Ridge. The annual horizon offers the strongest gains, consistent with the well-documented finding that return predictability is more pronounced at lower frequencies \citep{cochrane2011presidential}.
At the monthly frequency (panel~b), the noise-regularized $R^2$ is $+0.37\%$ at $p_{cv} = 12{,}000$, while CV-Lasso ($-0.46\%$) and CV-Ridge ($-1.59\%$) again fail to beat the historical mean. Although the monthly $R^2$ is modest in absolute terms, a positive out-of-sample $R^2$ at the monthly horizon is economically meaningful: as \cite{campbell2008predicting} note, monthly $R^2$ values in the range of 0.5\% can generate substantial utility gains for a mean-variance investor. The pattern is consistent at the quarterly frequency as well (Appendix~\ref{app:quarterly}). In all cases, the $R^2$ for PCA is strongly negative (see Table~\ref{tab:summary}).

Three features merit emphasis. First, the cross-validated $p_{cv}$ scales roughly in proportion to the sample size: $p_{cv}/n$ is approximately 50 at all three frequencies ($1{,}000/20$, $4{,}000/80$, $12{,}000/240$), suggesting a stable optimal noise-to-signal ratio. Second, the $R^2$ curve is remarkably flat beyond $p_{cv}$; even doubling the number of noise variables has little effect, confirming the robustness documented in the simulations. Third, the improvement over CV-Lasso and CV-Ridge does not come from better variable selection but from the opposite direction: by adding pure noise, the method achieves implicit regularization that is better adapted to the dense, weak-factor structure of equity premium predictors than explicit penalization with a single tuning parameter.

\section{Conclusion}
\label{sec:conclusion}
When economic outcomes are driven by latent factors shared across a large predictor panel, the reduced-form forecasting model is dense. This paper takes denseness as a starting point and establishes four results. The population working model converges to the oracle that uses the true factors as predictor dimension grows (Theorem~\ref{th1}). Perfect variable selection fails to achieve first-order forecast optimality when the number of retained predictors is comparable to $n$, because the design matrix lacks the dimensionality needed for eigenvalue regularization (Proposition~\ref{th:3ols}). Noise regularization, i.e., retaining or deliberately adding uninformative predictors restores optimality by inflating the eigenvalue spectrum, without estimating the factors or their number (Theorems~\ref{th2} and~\ref{th4denoise}). The AR($p$) symmetry property clarifies that overparameterization through additional lags cannot substitute for expanding the predictor set (Proposition~\ref{th:ar}).

Our empirical applications illustrate this principle in three areas of economic forecasting. Noise regularization improves and stabilizes forecasts of U.S.\ inflation, cross-country GDP growth, and the U.S.\ equity risk premium, outperforming Lasso, Ridge, and PCA benchmarks.  The central message is that the relevant margin for improving dense linear forecasts is the dimension of the predictor space, not the informativeness of its individual components.

\bigskip

\appendix

\section{Proofs}

This appendix elaborates technical proofs of the theoretical statements in the main text.

\subsection{Proof of Theorem \ref{th1}}
% ZT checked. 2026-2-23

\begin{proof}

The factor model implies that $\Sigma_{X}=\Lambda\Sigma_{f}\Lambda'+\Sigma_{u}$,
where $\Sigma_{f}=\mathbb{E}f_{t}f_{t}'$ is invertible.
By the definition of linear projection, the projection coefficient is
\[
\beta=[ \mathbb E x_tx_t']^{-1}\mathbb{E}x_{t}y_{t}=\Sigma_{X}^{-1}\mathbb{E}\left(\Lambda f_{t}+u_{t}\right)(f_{t}'\rho+\epsilon_{y,t})=\Sigma_{X}^{-1}\Lambda\Sigma_{f}\rho
\]
under the setting of \eqref{eq2.1}  and \eqref{eq2.2} with $\mathbb{E}\left[f_{t}  (u'_{t},\epsilon_{y,t})\right]=0$.

\textbf{Part (i)}.   By the Woodbury matrix identity, we expand $\beta$ as 
\begin{align*}\label{eq:beta-explicit}
\beta & = 
\Sigma_{X}^{-1}\Lambda\Sigma_{f}\rho \nonumber   =\left(\Sigma_{u}^{-1}-\Sigma_{u}^{-1}\Lambda(\Sigma_{f}^{-1}+\Lambda'\Sigma_{u}^{-1}\Lambda)^{-1}\Lambda'\Sigma_{u}^{-1}\right)\Lambda\Sigma_{f}\rho \nonumber \\
 & =\Sigma_{u}^{-1}\Lambda\left(I-(\Sigma_{f}^{-1}+\Lambda'\Sigma_{u}^{-1}\Lambda)^{-1}\Lambda'\Sigma_{u}^{-1}\Lambda\right)\Sigma_{f}\rho \nonumber   =\Sigma_{u}^{-1}\Lambda(\Sigma_{f}^{-1}+\Lambda'\Sigma_{u}^{-1}\Lambda)^{-1}\rho.
\end{align*}
The $L_{2}$-norm of $\beta$ is bounded by
\begin{align*}
\left\Vert  \beta \right\Vert  & \leq\left\Vert \Sigma_{u}^{-1/2}\right\Vert \left\Vert \Sigma_{u}^{-1/2}\Lambda(\Sigma_{f}^{-1}+\Lambda'\Sigma_{u}^{-1}\Lambda)^{-1}\right\Vert \left\Vert \rho\right\Vert   \leq\left\Vert \Sigma_{u}^{-1/2}\right\Vert \left\Vert (\Sigma_{f}^{-1}+\Lambda'\Sigma_{u}^{-1}\Lambda)^{-1/2}\right\Vert \left\Vert \rho\right\Vert \\
 & \leq\left\Vert \Sigma_{u}^{-1/2}\right\Vert \left\Vert (\Lambda'\Sigma_{u}^{-1}\Lambda)^{-1/2}\right\Vert \left\Vert \rho\right\Vert   =\frac{\Vert \Sigma_{u}^{-1/2}\Vert }{\sigma_{\min}(\Sigma_{u}^{-1/2})}\left\Vert (\Lambda'\Lambda)^{-1/2}\right\Vert \left\Vert \rho\right\Vert.
\end{align*}
Under Assumption \ref{ass:Xt}, we have $ \| \Lambda' \Lambda \| \asymp \psi_{p,n}$, while 
$$
\sigma_{\max}(\Sigma_{u})/\sigma_{\min}(\Sigma_{u}) = C_U / c_u \in(0,\infty)
$$
and 
$\left\Vert \rho\right\Vert =O(1)$ do not change the order.
As a result, $\left\Vert  \beta \right\Vert = O\left(\psi_{p,n}^{-1/2}\right). $

\textbf{Part (ii)}. We decompose the error term of the working model as 
$$
e_t =y_t -x_t '\beta=\epsilon_{y,t }+f_t '(\rho-\Lambda'\beta)-u_t '\beta.
$$
It follows
$$
\mathbb{E}[(u_t '\beta)^{2}]=\beta'\Sigma_{u}\beta\leq C_U \left\Vert \beta\right\Vert ^{2}=O\left(\psi_{p,n}^{-1}\right).
$$
Moreover, the explicit expression of $\beta$ yields
\[
\rho-\Lambda'\beta=\rho-\Lambda'\Sigma_{X}^{-1}\Lambda\Sigma_{f}\rho=\left(I_{K}-\Lambda'\left(\Lambda\Sigma_{f}\Lambda'+\Sigma_{u}\right)^{-1}\Lambda\Sigma_{f}\right)\rho=\Sigma_{f}^{-1}(\Sigma_{f}^{-1}+\Lambda'\Sigma_{u}\Lambda)^{-1}\rho
\]
by the Woodbury matrix identity, so that its norm is bounded by 
\begin{align}\label{eq:rho-Lambda_beta}
\left\Vert \rho-\Lambda'\beta\right\Vert & \leq \left\Vert \Sigma_{f}^{-1}\right\Vert \left\Vert (\Sigma_{f}^{-1}+\Lambda'\Sigma_{u}\Lambda)^{-1}\right\Vert \left\Vert \rho\right\Vert  \\ \nonumber
& \leq \left\Vert \Sigma_{f}^{-1}\right\Vert \left\Vert (\Lambda'\Sigma_{u}\Lambda)^{-1}\right\Vert \left\Vert \rho\right\Vert =O\left(\psi_{p,n}^{-1}\right).  
\end{align}
Hence
$
\Var(f_t '(\rho-\Lambda'\beta))\leq \|\rho-\Lambda'\beta\|^2 \, \tr ( \Sigma_f ) = O(\psi_{p,n}^{-2}).
$

The conclusion follows given that $\mathrm{Var}\left(f_{t}'(\rho-\Lambda'\beta)-u_{t}'\beta\right)=O\left(\psi_{p,n}^{-1}\right)$.
\end{proof}

\subsection{Auxiliary lemmas}

The proof of Theorem \ref{th1} is calculated in the population. To proceed to the sample, we introduce a few auxiliary lemmas in this standalone section.  These lemmas will be used repeatedly in the remaining theoretical results. The proofs are given in the Supplement.

\begin{lemma} \label{lem:von_neumann}
Let $A$ be positive semidefinite and $B$ be a generic compatible matrix. Then 
\begin{enumerate}[label=(\roman*).]
\item $\tr(AB)\leq \|B\|\tr(A). $
\item $\tr(B'AB)\geq\lambda_{\min}(A)\tr(BB')$ 
\end{enumerate}
\end{lemma}

The following lemma helps clarify the relationship between conditional
expectation and stochastic orders.
%It will be used in the proof of Theorem 2.

\begin{lemma} \label{lem:cond-order} For four generic random variables
$a,b,d\geq0$ and $X$, if $\mathbb{E}\left[a|X\right]\leq b\mathbb{E}\left[d|X\right]$
with $b=O_{P}(r_{n})$ and $\mathbb{E}\left[d\right]=\mu_{d}\in(0,\infty)$,
then $\mathbb{E}\left[a|X\right]=O_{P}(r_{n})$.
\end{lemma}

\bigskip

The above two lemmas use generic notations. 
In contrast, the following lemma will involve random matrices in the context of our paper. 
Recall $B_{X}=(X'X)^{+}X'$. 

\begin{lemma} \label{lem:XX_orders} 
Suppose Assumptions \ref{ass:Xt}  and \ref{ass:e} hold. When $n,p\to\infty$ and $n/p\to0$, we have
\begin{enumerate}[label=(\roman*).]
\item 
$\frac{9}{16}c_u\leq\sigma_{\min}\left(UU'/p\right)\leq\sigma_{\max}\left(UU'/p\right)\leq  \frac{25}{16} C_u$
w.p.a.1.

\item $\sigma_{\min}\left(XX'/p\right)\geq\frac{1}{4}c_u$ w.p.a.1.

\item $\|B_X\|^2=\|(X'X)^+\|=O_P(1/p)$ and $\|B_{X}'\Lambda\|^{2}=O_{P}\left(1/n\right)$.
\end{enumerate}
\end{lemma}

\subsection{Proof of Proposition \ref{th:3ols}}

\begin{proof}

Since we work with the informative regressors only, the closed-form solution of the Ridgeless with
the regressors in the set ``$I$'' is 
$\widetilde{\beta}_{I} =(X_{I}'X_{I})^{+}X_{I}'y.
$
Given that this is a negative message about the Ridgeless regression with $p_0$ regressors, it suffices to consider a special case where $\epsilon = (\epsilon_t)_{t=1}^n$ is independent over $t$ and independent of $U$ and $F$. The independence leads to the decomposition 
$$
 \mathbb{E}[(y_{\text{new}}-\widetilde{y}_{\new}^I)^2 | X_I ]
 = \sigma^2_{\epsilon}  + \mathbb{E}[(\rho'f_{\text{new}}-\widetilde{y}_{\new}^I)^2 | X_I ].
$$
Notice $ \mathbb  E [x_{I, \text{new}} x_{I,\text{new}}'| X_I] 
=   \Lambda_I \mathbb E(f_{\new}f_{\new}'|X_I)\Lambda_I' + \Sigma_{u,I} $ % due to the independence between the training data and the test data, 
under Assumption \ref{ass:Xt} (ii), we have   
\begin{align*}
 \mathbb{E}[(\rho'f_{\text{new}}-\widetilde{y}_{\new}^I)^2 | X_I ] & =\sigma_{\epsilon}^{2}\,\tr\left[ \mathbb  E [x_{I, \text{new}} x_{I,\text{new}}'| X_I]  (X_I'X_I)^{+}\right]  \geq\sigma_{\epsilon}^{2}\,\tr\left[ \Sigma_{u,I} (X_I'X_I)^{+}\right]  \\
& \geq  \sigma_{\min}(\Sigma_{u,I})  \sigma_{\epsilon}^{2}\, \tr \left[(X_I'X_I)^{+}\right]
\geq c_u \sigma_{\epsilon}^{2}\, \tr \left[(X_I'X_I)^{+}\right]
\end{align*}
by Lemma \ref{lem:von_neumann} (ii). 
We focus on the trace 
\begin{align*}
    \tr \left[(X_I'X_I)^{+}\right] & = \sum_{j=1}^{p_0\wedge n} \sigma^{-1}_j(X_I' X_I)  
    > \sum_{j=K+1}^{p_0\wedge n} \sigma^{-1}_j(X_I' X_I)   \geq \frac{p_0 \wedge n}{n} \sigma^{-1}_{K+1} \left(\frac{X_I'X_I}{n}\right),
\end{align*}
where the strict inequality holds when we remove  the top $K$ eigenvalues from the summation. 
When $p_0/n\to \gamma_0 \in (0,1)$, \citet[Theorem 2.5]{cai2020limiting} shows the $(K+1)$-th sample eigenvalue --- the largest non-spike eigenvalue 
$\sigma_{K+1} (X_I'X_I/ n)$ --- is a consistent estimator for the corresponding population eigenvalue $\sigma_{K+1}(\Sigma_{X,I}) > 0$.
There exists a constant $c$ such that 
$$
 \mathbb{E}[(\rho'f_{\text{new}}-\widetilde{y}_{\new}^I)^2 | X_I ] > c (\gamma_0 \wedge 1) 
$$
bounded away from 0 when the sample size is sufficiently large. Therefore, we conclude that there is an asymptotically non-vanishing gap between 
$ \mathbb{E}[(y_{\text{new}}-\widetilde{y}_{\new}^I)^2 | X_I ]$ and $\sigma^2_{\epsilon} $. The conclusion follows.
\end{proof}

% Proof of Proposition \ref{th:ar} is a linear algebraic fact and is relegated to Supplementary Appendix.

\subsection*{Proof of Proposition \ref{th:ar}}

\begin{proof}
  For an AR$(p)$ without an intercept, the $(T-p)\times p$ regressor matrix $X$, the $T-p$ outcome vector $y$, and the $p$ dimensional ``new" regressor $x_{\new}$ are respectively given as
$$
\underset{(T-p)\times p}{X}= \begin{pmatrix}
    y_{1:T-p}, y_{2:T-p+1},\cdots, y_{p:T-1}
\end{pmatrix}, \quad y= y_{p+1:T}, \, \mbox{ and }\, x_{\new}=y_{T-p+1:T}.
$$
Meanwhile, for the AR$(T-p)$ model, the regressor matrix $\widetilde X$ is of dimension $p\times (T-p)$, the outcome vector $\widetilde y$ is of dimension $p$, and the ``new'' regressor $\widetilde x_{\new}$ is $T-p$ dimensional, given respectively as:
$$
\widetilde X = X',\quad\widetilde y= x_{\new},\, \mbox{ and }\, \widetilde x_{\new} = y.
$$
If $p> T/2$, the predicted value is 
\begin{eqnarray*}
    \widehat y^{AR}_{T+1}(p) &=&x_{\new}' (X'X)^+X' y = x_{\new}' X' (XX')^{-1}y= \widetilde y'\widetilde 
     X(\widetilde X '\widetilde X )^{-1}\widetilde x_{\new}
    \cr 
    &=& \widetilde x_{\new}'(\widetilde X '\widetilde X )^{-1}\widetilde 
     X'\widetilde y = \widehat y^{AR}_{T+1}(T-p) 
\end{eqnarray*}
By symmetry, 
the case of $p\leq T/2$ maintains the equivalence. 
\end{proof}

\subsection{Proof of Theorem \ref{th2}}

\begin{proof}
 
% We are interested in $\mathbb E[(x_{\new}'(\widehat{\beta}-\beta))^2|X]$.

Conditioning on $X$, the predictive MSE is 
$$
\mathbb E[(y_{new}-\widehat y_{\new})^2|X]
=\Var(\epsilon_{y,\new}|X)+\mathbb E\left[(\rho'f_{\new}-  \widehat {y}_{\new} )^2|X\right].
$$
The first term is the optimal forecast MSE using the true latent factors, which cannot be reduced by any estimation method. 
The second term, which depends on the estimator,  can be decomposed as
\begin{align} \label{eq:decomp1}
\rho'f_{\new}-\widehat{y}_{\new}
 =\left(\rho'f_{\new}-x_{\new}'\beta\right)-x_{\new}'(\widehat{\beta}-\beta),
\end{align}
where   $ \rho'f_{\new}-x_{\new}'\beta $
is the model approximation error due to the working linear model
\eqref{eq3.2},
which is exactly the gap between $\epsilon_{y,\new}$ and $e_{\new}$.  
The second term $x_{\new}'(\widehat{\beta}-\beta)$ raises from the estimation error of $\hat{\beta}$, which can be further
decomposed into 
\begin{align}\label{eq:decomp2}
    x_{\new}'(\widehat{\beta}-\beta)
    =\left\{ x_{\new}'A_{X}\beta+x_{\new}'B_{X}\left(F(\rho-\Lambda'\beta)-U\beta\right)\right\} +x_{\new}'B_{X}\epsilon_{y},
\end{align}
where we recall $A_X= (X'X)^+X'X-I_p$ and $B_X= (X'X)^{+}X'$.

%We focus on the decomposition of the prediction error \eqref{eq:decomp2}
%due to the estimation error $(\widehat{\beta} - \beta)$, where the three terms on the right-hand side are:
%\begin{enumerate}
 %   \item $x_{\new}'B_{X}\epsilon_{y}$ associated with the variance, since under Assumption \ref{ass:e} (ii) it is uncorrelated with the other two terms once it is conditional on $(x_{\new},X)$;
   % \item $x_{\new}'A_{X}\beta$ associated with the bias arising from the $p>n$ regime. When $p<n$, this term is zero as $A_X=0$; 
   % \item $x_{\new}'B_{X}\left(F(\rho-\Lambda'\beta)-U\beta\right)$ associated with the approximation error of the working model. Under a conditional linear model  assumption $\mathbb E[y | X] = X'\beta$, we do not need to consider this approximation error. However, we prefer avoiding such a strong assumption and thus must take the approximation error into consideration.
%\end{enumerate}

We start with the \textbf{variance}
$
\mathbb E\left\{x_{\new}'B_X\epsilon_y\epsilon_y'B_X'x_{\new} |X\right\}.
$
By the law of iterated expectations, 
\begin{align*}
\mathbb{E}\left\{ x_{\new}'B_{X}\epsilon_{y}\epsilon_{y}'B_{X}'x_{\new}|X\right\}  & =\tr\left(\mathbb{E}\left\{ \mathbb{E}\left[x_{\new}'B_{X}\epsilon_{y}\epsilon_{y}'B_{X}'x_{\new}|X,x_{\new}\right]|X\right\} \right)\\
 & \leq\tr\left(\mathbb{E}\left\{ B_{X}'\mathbb{E}\left[x_{\new}x_{\new}'\,|\,X\right]B_{X}\,|\,X\right\} \right)\left\Vert \mathbb{E}\left[\epsilon_{y}\epsilon_{y}'|X, x_{\new}\right]\right\Vert \\
 & \leq \sigma^2_{\epsilon} \tr\left((X'X)^{+} \Sigma_{X_{\new}} \right)
\end{align*}
where $\Sigma_{X_{\new}} = \mathbb E(x_{\new}x_{\new}'\,|\,X).$
Using Lemma \ref{lem:von_neumann} (i), 
\begin{eqnarray*}
\tr\left((X'X)^{+} \Sigma_{X_{\new}} \right) & = & \tr\left((X'X)^{+}\Lambda \mathbb E\left[f_{\new}f_{\new}'|X\right]\Lambda'\right)+\tr\left((X'X)^{+}\mathbb{E}\left[u_{\new}u_{\new}'|X\right]\right)\\
& \leq  & C_f\tr\left((X'X)^{+}\Lambda \Lambda'\right)+ C_U\tr\left((X'X)^{+}\right)
\end{eqnarray*}
because by Jensen's inequality
\begin{align}\label{eq:Cf}
\Vert \mathbb E\left[f_{\new}f_{\new}'|X\right] \Vert  & 
\leq  \mathbb E\left[ \Vert \mathbb{E}[ f_{\new}f_{\new}'|F,U]\Vert \, |X\ \right]
=  \mathbb E\left[ \Vert \mathbb{E}[ f_{\new}f_{\new}'|F]\Vert \, |X\ \right]
\leq C_f
\end{align}
where the equality holds due to the independence between $f_{\new}$ and $U$ implied by Assumption \ref{ass:e} (i) and the bound by Assumption \ref{ass:e} (iii), and 
$$
\Vert \mathbb{E}\left[u_{\new}u_{\new}'|X\right] \Vert = 
\Vert \mathbb{E}\left[u_{\new}u_{\new}'\right] \Vert = \Vert \Sigma_u \Vert \leq C_U
$$ under Assumption \ref{ass:e} (i) and the bound by Assumption \ref{ass:Xt} (ii). 
Lemma \ref{lem:XX_orders} (ii) and (iii) give
\begin{eqnarray*}
\tr\left((X'X)^{+}\Lambda\Lambda'\right)&=&\tr(\Lambda'B_{X}B_{X}'\Lambda)\leq K\left\Vert B_{X}'\Lambda\right\Vert ^{2}=O_{P}(1/n);
\cr 
\tr\left((X'X)^{+}\right)&=&\sum_{j=1}^{n}\sigma_{j}((XX')^{-1})\leq\frac{n}{\sigma_{\min}\left(XX'\right)}=O_{P}(n/p).
\end{eqnarray*}
The order of the variance is 
\begin{equation}
\mathbb E\left\{x_{\new}'B_X\epsilon_y\epsilon_y'B_X'x_{\new} |X\right\} =O_{P}\left(\frac{1}{n}+\frac{n}{p}\right)\label{eq:var_main_part}.
\end{equation}
% \bigskip

Next, we move on to the other two terms in the decomposition \eqref{eq:decomp2}.
We first work with the  {squared bias} $\mathbb E\left\{\left[x_{\new}'A_{X}\beta \right]^2|X\right\}$.
Using the notations laid out in the main text, we apply a singular
decomposition $X=U_{n}S_{n}V_{n}'$ (notice $V_{n}$ is a $p\times n$
matrix, whose columns are the singular vectors of $X$ corresponding
to the first $n$ singular values), and thus 
\[
X'X=V_{n}S_{n}^{2}V_{n}',\quad(X'X)^{+}=V_{n}S_{n}^{-2}V_{n}'
\] where $V_{-n}$ is a
$p\times\left(p-n\right)$ matrix, columns being eigenvectors of the
$p\times p$ matrix $X'X$ corresponding to the $p-n$ eigenvalues. Obviously, $V_{-n}'V_{n}=0$. 
We rewrite 
\begin{align*}
\mathbb E\left\{\left[x_{\new}'A_{X}\beta \right]^2|X\right\} 
&=\beta'A_{X} \Sigma_{X_{\new}} A_{X}\beta 
=\beta'V_{-n}V_{-n}'\Sigma_{X_{\new}} V_{-n}V_{-n}'\beta\\
&\leq\left\Vert V_{-n}'\Sigma_{X_{\new}} V_{-n}\right\Vert \left\Vert V_{-n}\right\Vert ^{2}\left\Vert \beta\right\Vert ^{2} \leq\left\Vert V_{-n}'\Sigma_{X_{\new}}V_{-n}\right\Vert O(\psi_{p,n}^{-1})
\end{align*}
where and $A_{X}=V_{n}V_{n}'-I_{p}=-V_{-n}V_{-n}'$. 
Again by \eqref{eq:Cf}:
\begin{align*}
\left\Vert V_{-n}'\Sigma_{X_{\new}} V_{-n}\right\Vert  & \leq\left\Vert V_{-n}'\Lambda\mathbb{E}(f_{\new}f_{\new}'|X)\Lambda'V_{-n}\right\Vert +\left\Vert V_{-n}'\mathbb{E}(u_{\new}u_{\new}'|X)V_{-n}\right\Vert \\
 & \leq\left\Vert V_{-n}'\Lambda\right\Vert ^{2}\left\Vert \mathbb{E}(f_{\new}f_{\new}'|X)\right\Vert +\left\Vert \mathbb{E}(u_{\new}u_{\new}'|X)\right\Vert \left\Vert V_{-n}\right\Vert ^{2}\\
 & \leq C_f\left\Vert V_{-n}'\Lambda\right\Vert ^{2}+C_U.
\end{align*}
To deal with $V_{-n}'\Lambda$, let $V_{K}$ denote a $p\times K$ matrix whose
columns are the eigenvectors of $X'X$ corresponding to the first
$K$ eigenvalues. Two useful facts follow from standard theories of factor
models:

\begin{description}
    \item[Fact 1] The top $K$ eigenvalues of $X'X/n$ grow at rate $\psi_{p,n}$ as long as $\psi_{p,n}\gg p/n$;

    \item[Fact 2] There exists an $K\times K$ matrix $H$ such that $\|H^{-1}\|=O_{P}(\psi_{p,n}^{1/2})$
and 
\[
\|V_{K}-\Lambda H\| =  O_{P}\left(\sqrt{p / (\psi_{p,n}n)}\right).
\]
\end{description}

We decompose 
$
V_{-n}'\Lambda=V_{-n}'\left(\Lambda H-V_{K}+V_{K}\right)H^{-1}=V_{-n}'\left(\Lambda H-V_{K}\right)H^{-1},
$
 where the second equality is due to the orthogonality $V_{-n}'V_{K}=0$.
Therefore, 
\begin{equation}
\|V_{-n}'\Lambda\|\leq  \|V_{-n}'(\Lambda H-V_K)\|\|H^{-1}\|\leq O_{P}\left(\sqrt{\frac{p}{\psi_{p,n}n}}\right)O_{P}(\sqrt{\psi_{p,n}})=O_{P}\left(\sqrt{\frac{p}{n}}\right).\label{eqvlamb}
\end{equation}
We conclude $$\beta'A_{X}\Sigma_{X_{\new}}A_{X}\beta\leq\left(C_f\left\Vert V_{-n}'\Lambda\right\Vert ^{2}+C_U\right)O(\psi_{p,n}^{-1})=O_{P}\left(\frac{p}{n\psi_{p,n}}\right).$$

Next, we deal with the two terms in the {approximation error}. We first work with
\begin{eqnarray}
  \mathbb E \{ [ x_{\new}'B_{X}F(\rho-\Lambda'\beta) ]^2|X \} 
 & = & \mathbb{E}\left\{ (\rho-\Lambda'\beta)'F'B_{X}' \Sigma_{X_{\new}} B_{X}F(\rho-\Lambda'\beta)|X\right\} \nonumber \\
 & \leq & \| \Sigma_{X_{\new}} \| \, \mathbb E \left\{    \| (\rho-\Lambda'\beta)'F'B_{X}'\|^2\, |\, X\right\}\label{eq:x_newBF1}
\end{eqnarray}
by the law of iterated expectations. 
Assumption \ref{ass:e} (i) and (iii) imply 
\begin{align}
\| \Sigma_{X_{\new}} \|  & =  
\left\Vert \mathbb{E}\left[\Lambda f_{\new}f_{\new}'\Lambda'+u_{\new}u_{\new}'|F\right]\right\Vert 
\nonumber \\
& \leq\left\Vert \Lambda\mathbb{E}\left[f_{\new}f_{\new}'|F\right]\Lambda'\right\Vert +\left\Vert \mathbb{E}\left[u_{\new}u_{\new}'|F\right]\right\Vert \nonumber \\
 & \leq\tr\left(\Lambda'\Lambda\right)C_f+C_U
 \leq O_P( \psi_{p,n}).\label{eq:x_newBF2}
\end{align}
Notice 
\begin{align*}
\mathbb{E}\left\{\| (\rho-\Lambda'\beta)'F'B_{X}'\|^2|X\right\}  & 
  \leq\left\Vert \rho-\Lambda'\beta\right\Vert ^{2}\left\Vert B_{X}'B_{X}\right\Vert \mathbb{E}\left\{ \tr\left(FF'\right)|X\right\} \\
 & =O_{P}\left(\frac{n}{\psi_{p,n}^{2}p} \right)\mathbb{E}\left\{ \tr\left(\frac{F'F}{n}\right)|X\right\} 
\end{align*}
given $\left\Vert \rho-\Lambda'\beta\right\Vert =O\left(\psi_{p,n}^{-1}\right)$
in (\ref{eq:rho-Lambda_beta}) and $\left\Vert B_{X}'B_{X}\right\Vert  =O_{P}\left(1/p\right)$
by Lemma \ref{lem:XX_orders} (iii). Because the unconditional
mean $\mathbb{E}\left[F'F/n\right]=\Sigma_{f}$ has all eigenvalues
bounded away from 0 and $\infty$, we invoke Lemma \ref{lem:cond-order}
to obtain 
\begin{align}
\mathbb{E}\left\{ (\rho-\Lambda'\beta)'F'B_{X}'B_{X}F(\rho-\Lambda'\beta)|X\right\}  &
= \left(\frac{n}{\psi_{p,n}^{2}p} \right).\label{eq:x_newBF3}
\end{align}
 Insert \eqref{eq:x_newBF2} and \eqref{eq:x_newBF3} into \eqref{eq:x_newBF1},
we have 
\[
 \mathbb E\left\{\left[ x_{\new}'B_{X}F(\rho-\Lambda'\beta)\right]^2 \, |\, X\right\} =O_{P}\left(\frac{n}{\psi_{p,n}p}\right).
\]

The  second term in the approximation error  is $x_{\new}'B_{X}U\beta$. By parallel arguments as for \eqref{eq:x_newBF1}, 
% \textbf{Bias $ \mathbb E\left\{\left[ x_{\new}'B_{X}U\beta\right]^2|X\right\}$.}
we bound its expected square by the law of iterated expectations:
\begin{align*}
\mathbb{E}\left[\beta'U'B_{X}'x_{\new}x_{\new}'B_{X}U\beta|X\right] & 
\leq \| \Sigma_{X_{\new}} \| \Vert 
\mathbb{E}\left[ \beta'U'B_{X}'B_{X}U\beta\, |\,X\right] \leq  O_P(\psi_{p,n})
\mathbb{E}\left[ \beta'U'B_{X}'B_{X}U\beta\, |\,X\right],
\end{align*}
where the last inequality follows by \eqref{eq:x_newBF2}. 
We use Lemma \ref{lem:von_neumann} (i) to bound  
\begin{align*}
\mathbb{E}\left[\beta'U'B_{X}'B_{X}U\beta|X\right] & 
\leq \left\Vert (X'X)^{+}\right\Vert  \mathbb{E}\left[\tr\left(U\beta\beta'U'\right)|X\right]  =\frac{n}{p}  \left\Vert \left(\frac{X'X}{p} \right)^{+}\right\Vert \mathbb{E}\left[  \beta'(U'U/n)\beta | X \right]\\
 & = O_{P}\left( \frac{n}{p} \right) \cdot O_P(\|\beta\|^2)  
 =O_{P}\left(\frac{n}{p \psi_{p,n}}  \right)
\end{align*}
by Assumption \ref{ass:e} (i) about $U$,
$\Vert ( X'X / p)^{+} \Vert =O_{P}\left(1\right)$
by Lemma \ref{lem:XX_orders} (iii), and $\left\Vert \beta\right\Vert =O (\psi_{p,n}^{-1/2} )$
from Theorem \ref{th1}. 
The above two display expressions give
\[
\mathbb{E}\left[\beta'U'B_{X}'x_{\new}x_{\new}'B_{X}U\beta|X\right]\leq O_P(\psi_{p,n})O_{P}\left(\frac{n}{p}\psi_{p,n}^{-1}\right)=O_P\left(\frac{n}{p}\right).
\]

% In summary,
% $$
% \bias^2\leq O_{P}\left(\frac{p}{n\psi_{p,n}}+\frac{n}{p}\right).
% $$
%  \bigskip

So far, we have handled all terms in the expected square of $x_{\new}'(\widehat{\beta}-\beta)$ as decomposed in \eqref{eq:decomp2}. In \eqref{eq:decomp1} there is an additional term $\left(\rho'f_{\new}-x_{\new}'\beta\right)$, due to the  {population approximation error} of the linear working model \eqref{eq3.2}, which is irrelevant to the method of estimation. The expected square this term is bounded by   
\begin{align*}
\mathbb{E}[(\rho'f_{\new}-x_{\new}'\beta)^{2}|X] & =\mathbb{E}[(f_{\new}'(\rho-\Lambda'\beta)-u_{\new}'\beta)^{2}|X]\\
 & \leq2\left\Vert \rho-\Lambda'\beta\right\Vert ^{2}\left\Vert \mathbb{E}[(f_{\new}f_{\new}'|X]\right\Vert +2\left\Vert \beta\right\Vert ^{2}\left\Vert \mathbb{E}[(u_{\new}u_{\new}'|X]\right\Vert \\
 & \leq2 \left\{ C_f\left\Vert \rho-\Lambda'\beta\right\Vert ^{2}+ C_U\left\Vert \beta\right\Vert ^{2}\right\} =O_{P}(\psi_{p,n}^{-1})
\end{align*}
given \eqref{eq:Cf} and 
the order provided in Theorem \ref{th1}.

To summarize, in this proof we have shown that the order of the variance is $O_P(1/n+ n/p)$ and the bias from $A_X$ is $O_P(p/(n\psi_{p,n}))$. These two are the leading terms, and the approximation error of the working model to the true factor model is no larger than them. We therefore conclude the stochastic order. 
Given the relative magnitude of $p$, $n$ and $\psi_{p,n}$, both the leading terms are $o_P(1)$ asymptotically, and thus the consistency follows, 
as stated in  {Part (ii)} of the statement.
\end{proof}

% Proof of Theorem \ref{th4denoise} is similar to the above one and is relegated to the Supplementary Appendix. 

\subsection*{Proof of Theorem \ref{th4denoise}}

\begin{proof}
Write $X=(X_I, X_N)$. Here $X_N$ is noise so that  $X=F \Lambda' + (U_{I}, X_N)$ with  $\Lambda'=(\Lambda_I',0)$ and $\|X_N\|= O_P(p)$. Correspondingly, the true coefficient in the working model is  $\beta= (\beta_I',0)'$. 
The estimated coefficient can be written as 
$$
\widehat\beta_I = X_I'(XX')^{-1}y=B_{I,X}y,\quad \mbox{ where } B_{I,X}=X_I'(X_IX_I'+X_NX_N')^{-1}.
$$
Define a $p_0\times p$ selection matrix $Q_I= (I_{p_0},0)$, and then the error in prediction is  
\[
x_{\new,I}'(\widehat{\beta}_I-\beta_I)= x_{\new,I}'Q_IA_{X}\beta+x_{\new,I}'B_{I,X}\left(F(\rho-\Lambda_I'\beta_I)-U_I\beta_I\right) +x_{\new,I}'B_{I,X}\epsilon_{y},
\]
where we recall   $A_X= X'(XX')^{-1}X-I_{p}$ .  The first term is the bias, the second term is the approximation error, and the last term is associated with the variance. 

The proofs of the squared expectations of the first two terms resemble those in Theorem \ref{th2}, and thus omitted for brevity. They share the same rate of convergence 
$
O_{P}\left(\frac{p}{n\psi_{p,n}}+\frac{n}{p}\right)
$
as their counterparts, respectively, in Theorem \ref{th2}.

The variance, on the other hand, enjoys an improved rate of convergence, as detailed  below.  We work with the case $p_{0}>n$; parallel argument
follows for $p_{0}\leq n$. 
Under a finite $K$, the spiked eigenvalues
of $X_{I}'X_{I}$ is of order $O_P(n\psi_{p,n})$ whereas the remaining
$(n-K)$ non-spiked eigenvalues are $O_{P}\left(p_{0}\right)$ \citep{cai2020limiting}, which give 
\begin{align}
\tr\left(X_{I}'X_{I}\right) & =\sum_{j=1}^{n}\sigma_{j}\left(X_{I}'X_{I}\right)=O_{P}\left(Kn\psi_{p,n}+(n-K)p_{0}\right)=O_{P}\left( np_{0}\right)\label{eq:order_tr_XX}
\end{align}
under the condition  $\psi_{p,n} = O(p_0)$ in Assumption \ref{ass:Xt} (i).
%and it further implies
%\begin{equation}
%\tr\left((X_{I}'X_{I})^{2}\right)=\sum_{j=1}^{n}\sigma_{j}^{2}\left(X_{I}'X_{I}\right)=O_{P}\left(n^2\psi_{p,n}^{2}+(n-K)p_{0}^{2}\right)=O_{P}\left(n^2\psi_{p,n}^{2}+np_{0}^{2}\right).\label{eq:order_tr_XX2}
%\end{equation}
 
Given Assumption \ref{ass:e} (ii),
the variance is bounded by 
\begin{eqnarray}
 & &\mathbb{E}\left[x_{\new,I}'B_{I,X}\epsilon_{y}\epsilon_{y}'B_{I,X}'x_{\new,I}|X\right]  \nonumber   \leq \sigma_{\epsilon}^2 \ \tr
\left\{ \Sigma_{X_{\new}} B_{I,X}B_{I,X}'\right\} \cr 
 & \leq & \sigma_{\epsilon}^2 \left\Vert \mathbb{E}\left[f_{\new}f_{\new}|X\right]\right\Vert \tr\left\{ \Lambda_{I}'B_{I,X}B_{I,X}'\Lambda_{I}\right\} + \sigma_{\epsilon}^2 \left\Vert \mathbb{E}\left[u_{\new,I}u_{\new,I}'|X\right]\right\Vert \tr\left\{ B_{I,X}B_{I,X}'\right\} \nonumber \\
 &\leq &  \sigma_{\epsilon}^2 C_f \tr\left\{ \Lambda_{I}'B_{I,X}B_{I,X}'\Lambda_{I}\right\} + \sigma_{\epsilon}^2 C_U \tr\left\{ B_{I,X}B_{I,X}'\right\}  ,\label{eq:tr_X_new_B}
\end{eqnarray}
where the last inequality follows by \eqref{eq:Cf}.
The second term of the above expression is bounded, using Lemma \ref{lem:XX_orders} (ii) and (\ref{eq:order_tr_XX}), by
\begin{align}
\tr\left\{ B_{I,X}B_{I,X}'\right\}  & =\tr\left(\left(XX'\right)^{-2}X_{I}'X_{I}\right)
\leq  \Vert \left(XX'\right)^{-2}  \Vert \tr\left(X_{I}'X_{I}\right)\nonumber \\
 & =O_{P}(p^{-2})\cdot O_{P}\left(np_{0}\right)=O_{P}\left( n p_0/p^2\right). 
 \label{eq:tr_BXBX}
\end{align}

The derivation of the order of $\tr\left\{ \Lambda_{I}'B_{I,X}B_{I,X}'\Lambda_{I}\right\} $
is similar to the proof of Lemma \ref{lem:XX_orders} (iii). 
Notice $\Lambda= (\Lambda_I',0)'$ be $p \times K$, and  $B_{X,I}$ is  the first $p_0$ rows of $B_X = X'(XX')^{-1}$.
Given  $U=(U_I, X_N)$ which is $n\times p$.  Define $\Vert \cdot \Vert_F$ as the Frobenius norm of a matrix.  In view of $X=F \Lambda' +U$, we can write 
\begin{eqnarray}
   \tr\left\{ \Lambda_{I}'B_{I,X}B_{I,X}'\Lambda_{I}\right\}&=& \|\Lambda_I'B_{I,X}\|_F^2
=\|\Lambda'B_{X}\|_F^2
=\| (F'F)^{-1}F'(X-U)B_{X}\|_F^2 \cr 
&\leq &O_P(n^{-1})\|XB_X\| + O_P(n^{-1})\|UB_X\|= O_P(n^{-1})
\end{eqnarray}
where the last equality follows from the same proof of Lemma \ref{lem:XX_orders} (iii). Putting together these pieces, we conclude that the variance is of order  
$
 O_P \left(\frac{1}{n}+\frac{np_0}{p^2} \right)
$
for forecast based on the denoised estimator.
\end{proof}

\bigskip

% \singlespacing

\bibliographystyle{chicago}
\bibliography{liao}

%---------------------------------------------------------------%
% SUPPLEMENT
%---------------------------------------------------------------%
\clearpage
\setcounter{page}{1}
\renewcommand{\thepage}{A\arabic{page}}
\setcounter{footnote}{0}

\begin{center}
{\Large\bf Benign Overfitting in Economic Forecasting \\ via Noise Regularization}\\[0.3in]
{\large Yuan Liao\footnote{Department of Economics, University of Iowa, \texttt{yuan-liao@uiowa.edu}} \quad
Xinjie Ma\footnote{Tippie College of Business, University of Iowa, \texttt{xinjie-ma@uiowa.edu}} \quad
Andreas Neuhierl\footnote{Mitch Daniels School of Business, Purdue University, \texttt{aneuhier@purdue.edu}} \quad
Zhentao Shi\footnote{Department of Economics, The Chinese University of Hong Kong, \texttt{zhentao.shi@cuhk.edu.hk}}}\\[0.4in]
{\Large\bf Supplemental Appendix}
\end{center}

\setcounter{section}{0}
\renewcommand{\thesection}{A.\arabic{section}}
\setcounter{equation}{0}
\renewcommand{\theequation}{A.\arabic{equation}}
\setcounter{theorem}{0}
\renewcommand{\thetheorem}{A.\arabic{theorem}}
\setcounter{table}{0}
\renewcommand{\thetable}{A.\arabic{table}}
\setcounter{figure}{0}
\renewcommand{\thefigure}{A.\arabic{figure}}

\onehalfspacing

\section{Additional Proofs}
\subsection*{Proof of Lemma \ref{lem:von_neumann}}
\begin{proof}
\textbf{Part (i)}. \cite{von1937some}'s trace inequality gives
\begin{align*}%\label{eq:von_neumann}
    \tr(AB)\leq \sum_i\sigma_i(A)\sigma_i(B)
\end{align*}
Since $A$ is positive semidefinite, we continue the inequality:
$$
\sum_i\sigma_i(A)\sigma_i(B) \leq \sum_i\sigma_i(A)| \sigma_i(B)| 
\leq \|B\| \sum_i\sigma_i(A) = \|B\| \tr(A). 
$$

\textbf{Part (ii)}. Denote the dimension of $A$ as $m\times m $, and thus $BB' $ is also an $m\times m$ positive semidefinite matrix. There is another corollary of von Neumann's trace inequality: 
\begin{align*}
\tr(B'AB) & = 
\tr(A BB' )\geq \sum_{i=1}^{m}\sigma_i(A)\sigma_{m-i+1}(BB' ) \\
 & \geq 
\sigma_{\min} (A) \sum_{i=1}^{m}\sigma_{m-i+1}(BB' ) 
= \sigma_{\min} (A) \tr(BB' ).     \qedhere
\end{align*}
\end{proof}

\subsection*{Proof of Lemma \ref{lem:cond-order}}

\begin{proof}
This is an elementary and straightforward result. We lay out the proof
for completeness. For any constants $C_{b},C_{d}>0$ and  $C_{a}=2C_{b}C_{d}$, we have 
\begin{eqnarray*}
&& P\left(\mathbb{E}\left[a|X\right]\geq C_{a}r_{n}\right) \\
&\leq& P\left(\mathbb{E}\left[a|X\right]\geq C_{a}r_{n}\,|\,d\leq C_{d}\right)+P\left(d\geq C_{d}\right)\\ 
 &\leq& P\left(b\cdot\mathbb{E}\left[d|X\right]\geq2C_{b}C_{d}r_{n}\, |\, d\leq C_{d},b\leq C_{b}r_{n}\right)+P\left(b\geq C_{b}r_{n}\right)+P\left(d\geq C_{d}\right)\\ 
 &=&P\left(b\geq C_{b}r_{n}\right)+P\left(d\geq C_{d}\right)\\ &\leq& P\left(b\geq C_{b}r_{n}\right)+\frac{\mu_{d}}{C_{d}},
\end{eqnarray*}
where the last inequality follows from the Markov inequality.
 Under the assumption $b=O_{P}(r_{n})$ and $\mu_{d}\in(0,\infty)$,
we choose $C_{b}$ and $C_{d}$ sufficiently large to make the
right-hand side of the above inequality arbitrarily small. 
\end{proof}

\subsection*{Proof of Lemma \ref{lem:XX_orders}}

\begin{proof} 
\textbf{Part (i)}
For a constant $ \gamma \in(0,1/16)$, define
$n_{\gamma}=\left\lfloor \gamma p\right\rfloor $. 
Since $n/p \to 0$, without loss of generality, we can assume $p$ is sufficiently large so that $n_{\gamma} > n$. We define an $n_{\gamma}\times p$ matrix
$W^{*}= (W', \tilde{W}')'$ where $\tilde{W}$ is an $(n_{\gamma}-n)\times p$ matrix with i.i.d.~entries
that follow the same distribution as those in $W$. 
\citet[Theorem 2]{bai1993limit} yields
\[
\sigma_{\max}(W^{*}W^{*\prime}/p)\stackrel{a.s.}{\to} (1+\sqrt{\gamma})^2
\ \quad \text{and}\ \quad \sigma_{\min}(W^{*}W^{*\prime}/p)\stackrel{a.s.}{\to} (1-\sqrt{\gamma})^2
\]
as $n\to \infty$ and  $n_{\gamma}/p\to \gamma$.
Since $WW'$ is the top-left submatrix block of $W^{*}W^{*\prime}$, 
w.p.a.1 we have
\[
\frac{9}{16}\leq\sigma_{\min}\left( \frac{W^{*}W^{*\prime}}{p}\right)
\leq\sigma_{\min}\left( \frac{WW^{\prime}}{p}\right)
\leq\sigma_{\max}\left( \frac{WW^{\prime}}{p}\right)
\leq\sigma_{\max}\left( \frac{W^{*}W^{*\prime}}{p}\right)\leq\frac{25}{16}.
\]
It follows that w.p.a.1 the minimum eigenvalue
\[
\sigma_{\min}\left(UU'/p\right)\geq\sigma_{\min}(\Sigma_{u})\cdot\sigma_{\min}(WW'/p)\geq\frac{9}{16}c_u
\]
and the maximum eigenvalue
\[
\sigma_{\max}(UU'/p)\leq\left\Vert \Sigma_{u}\right\Vert \cdot\sigma_{\max}(WW'/p)\leq\frac{25}{16}C_u
\]
as stated. 

\textbf{Part (ii).} For any\textbf{ $v\in\mathbb{R}^{n}$ }such that
$\left\Vert v\right\Vert =1$, we decompose 
\begin{align*}
v'\frac{XX'}{p}v & =v'F\frac{\Lambda'\Lambda}{p}F'v+2v'F\frac{\Lambda'U'}{p}v+v'\frac{UU'}{p}v=\alpha'\alpha+2\alpha\eta+v'\frac{UU'}{p}v\\
 & \geq\|\alpha\|^{2}-2\|\alpha\|\|\eta\|+v'\frac{UU'}{p}v,
\end{align*}
where $\alpha=\left(\Lambda'\Lambda/p\right)^{1/2}F'v$ and $\eta=\left(\Lambda'\Lambda/p\right)^{-1/2}\frac{\Lambda'}{p}U'v$
are two $r$-dimensional vectors. The latter is bounded by
\begin{align*}
\|\eta\|^{2} & =v'U\frac{\Lambda}{p}\left(\frac{\Lambda'\Lambda}{p}\right)^{-1}\frac{\Lambda'}{p}U'v\\
 & \leq\frac{1}{p \sigma_{\min}(\Sigma_{u}) }\cdot v'W\Sigma_{u}^{1/2}\frac{\Lambda}{\sqrt{p}}\left(\frac{\Lambda'\Sigma_{u}\Lambda}{p}\right)^{-1}\frac{\Lambda'}{\sqrt{p}}\Sigma_{u}^{1/2}W'v\\
 & \leq (c_u p )^{-1} \sigma_{\max} (\tilde{W}'\tilde{W} )
\end{align*}
where $\tilde{W}=\left(\frac{\Lambda'\Sigma_{u}\Lambda}{p}\right)^{-1/2}\frac{\Lambda'}{\sqrt{p}}\Sigma_{u}^{1/2}W'$
is a $K\times n$ matrix with i.i.d.~entries of zero mean and unit variance.
Since $K$ is finite, 
we have $\frac{n}{p}\sigma_{\max}\left(\frac{\tilde{W}'\tilde{W}}{n}\right)\stackrel{p}{\to}\lim\frac{n}{p}=0,$
and therefore $\left\Vert \eta\right\Vert =o_{p}(1)$; in other words, w.p.a.1
the event $\left\{ \|\eta\|\leq\sqrt{c}\right\} $ occurs. Under
this event, and together with Part (i), we have 
\[
\sigma_{\min}\left(XX'/p\right)\geq\|\alpha\|^{2}-\left\Vert \alpha\right\Vert \sqrt{c_u}+\frac{9}{16}c_u = (\|\alpha\|-\frac{1}{2}\sqrt{c})^{2}+\frac{5}{16}c_u > \frac{1}{4}c_u.
\]

\textbf{Part (iii)}. Note that $B_{X}B_{X}'=\left(X'X\right)^{+}$, and then Part (ii) immediately yields 
$$\|B_X\|^2= \|(X'X)^+\|\leq \sigma_{\min}^{-1}(X'X)\leq 4c_u^{-1}p^{-1} = O_P(1/p).$$ 

For the other expression, given $\Lambda=(X-U)'F(F'F)^{-1}$ and $(F'F/n)^{-1}=O_{P}(1)$, we have
\begin{eqnarray*}
\|B_{X}'\Lambda\|^{2} & = & \left\Vert B_{X}'\left(X-U\right)'F(F'F)^{-1}\right\Vert _{}^{2}\\
 & \leq & \frac{1}{n}\left\Vert \left(X-U\right)B_{X}\right\Vert ^{2}\left\Vert \frac{F}{\sqrt{n}}\left(\frac{F'F}{n}\right)^{-1}\right\Vert ^{2}\\
 & \leq & \frac{2}{n}\left\{ \left\Vert XB_{X}\right\Vert ^{2}+\left\Vert UB_{X}\right\Vert ^{2}\right\} O_{P}(1)\\
 & = & \frac{2}{n}\left\{ 1+\left\Vert UB_{X}\right\Vert ^{2}\right\} O_{P}(1),
\end{eqnarray*}
where the last line follows by the fact that $XB_{X}=X\left(X'X\right)^{+}X'$
is idempotent, thereby $\left\Vert XB_{X}\right\Vert =1$. On the
other hand, w.p.a.1 we have 
\begin{align*}
\left\Vert UB_{X}\right\Vert ^{2} = \left\Vert U\left(X'X\right)^{+}U'\right\Vert ^{2}\leq\left\Vert \left(\frac{X'X}{p}\right)^{+}\right\Vert \left\Vert \frac{UU'}{p}\right\Vert =\frac{\sigma_{\max}\left(UU'/p\right)}{\sigma_{\min}\left(XX'/p\right)}\leq 3\frac{C_u}{c_u}
\end{align*}
as shown in Parts (i) and (ii). We conclude $$\|B_{X}'\Lambda\|^{2}\leq\frac{2}{n}\left(1+3\frac{C_{u}}{c_{u}}\right)O_{P}(1)=O_{P}\left(1/n\right)$$ as stated in Part (iii).
\end{proof}

\section{Additional Empirical Results and Implementation Details}

\subsection{Quarterly equity premium prediction} \label{app:quarterly}

This appendix reports the quarterly-frequency equity premium prediction that complements the annual and monthly results in Section~\ref{sec:equity_premium}. The setup is identical: 15 predictors from \cite{Goyal2008comprehensive}, 20-year (80-quarter) rolling windows, out-of-sample evaluation from 1960 to 2024, and $p$ selected by time-series cross-validation on the burn-in period.

\begin{figure}[h!]
\centering
\includegraphics[width=0.65\textwidth]{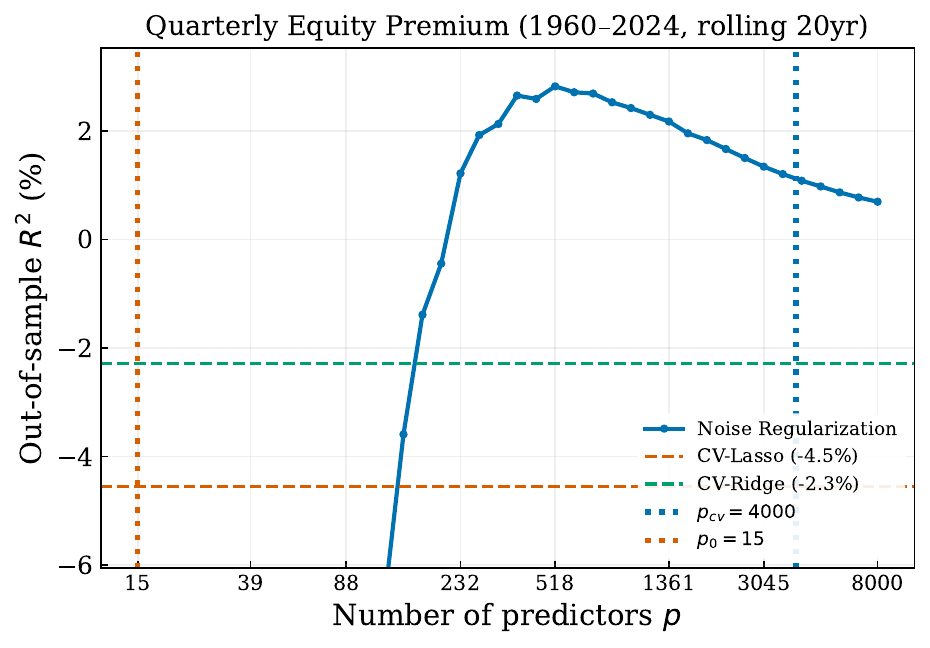}
\caption{Out-of-sample $R^2$ (\%) for predicting the U.S.\ equity premium at the quarterly frequency, 1960--2024, using 20-year rolling windows. The solid curve shows the noise-regularized forecast as $p$ increases. Horizontal dashed lines show CV-Lasso and CV-Ridge using only the 15 economic predictors. The vertical dotted lines mark $p_0 = 15$ and $p_{cv} = 4{,}000$.}
\label{fig:finance_quarterly_app}
\end{figure}

At the quarterly frequency, the noise-regularized forecast achieves $R^2_{\text{oos}} = +1.08\%$ at $p_{cv} = 4{,}000$, while CV-Lasso ($-4.55\%$) and CV-Ridge ($-2.29\%$) remain negative (Figure~\ref{fig:finance_quarterly_app}). The results are fully consistent with the annual and monthly patterns reported in the main text: the $R^2$ curve exhibits the characteristic second-descent shape, and $p_{cv}/n \approx 50$ aligns with the ratio observed at the other two frequencies.

\subsection{Cross Validation Algorithm} \label{sec:implementation}

\begin{algorithm}[Rao-Blackwell Leave-One-Out (LOO) for selecting $p$]\label{app:rbloo}
For each candidate total dimension $p$ on a grid $p_1 < \cdots < p_J$\textup{:}
\begin{enumerate}[nosep]
\item For $\ell = 1, \ldots, L$, draw $Z^{(\ell)} \sim N(0, I_{p - p_0})$ independently and form $X_{\mathrm{aug}}^{(\ell)} = [\mathbf{1}_n,\; X_I,\; Z^{(\ell)}]$. Compute
$$
e^{(\ell)}(p) = \Diag\!\bigl(G^{(\ell)}\bigr)^{-1} G^{(\ell)} Y, \qquad G^{(\ell)} = \bigl(X_{\mathrm{aug}}^{(\ell)} {X_{\mathrm{aug}}^{(\ell)}}'\bigr)^+.
$$
\item Average the LOO residual \emph{vectors} across draws\textup{:}
$$
e^{RB}(p) = \frac{1}{L} \sum_{\ell=1}^{L} e^{(\ell)}(p).
$$
\item Set $\mathrm{LOO}(p) = \| e^{RB}(p) \|^2$.
\end{enumerate}
Select $p_{cv} = \arg\min_{p} \mathrm{LOO}(p)$.
\end{algorithm}

\section{Simulation Comparisons with VAR  Forecast} \label{sec:var}

In Proposition \ref{th:ar} we have shown the algebraic fact of symmetry of AR($p$). The symmetry carries over into more general time series models. 
In this section, we conduct simulations to compare with the vector autoregressive model (VAR) with lags up to $L<T$. 

We generate an $N$-dimensional dynamic factor model:
$$
Y_{t+1}  =
\bar\Lambda f_{t+1} + E_{t+1},\quad \dim(Y_{t+1})= p_0,
$$
where $ Y_{t+1} = (Y_{j,t+1})_{j=1}^N $ is an $N$-dimensional vector and
the objective is to forecast the first time series $y_t=Y_{1,t+1}$. The  dynamic factors   satisfy a   VAR(1) model:
$$
f_{t+1}= \Theta_{f} f_{t}+ e_{f,t+1}.
$$ Let  $\bar\Lambda_{1}'$ denote the first row of $\bar\Lambda$, and $E_{1,t+1}$ denote the first element of $E_{t+1}$.  Then $ y_{t} = \rho'f_t + \epsilon_{y,t}, $  which takes the form of  (\ref{eq2.1}) with  $\rho =\Theta'_{f} \bar\Lambda_{1}$ 
and   $\epsilon_{y,t}=\bar\Lambda_1'e_{f,t+1} +E_{1,t+1}$. In addition, we use the one-period lagged variables as the ``informative predictors", which also satisfy a factor model:
$
X_{I,t} = Y_{t},
$ and merge them with additional added pure  noise (standard normal variables) to implement the Noise-regularization. 
We implement the  VAR($L$) model using Ridgeless regression, which is  overfitting when $L \times p_0> T-L$. Both methods can be compared in the same framework of Ridgeless regression: using the first lag of $Y_{t}$ as the main informative predictors, and additionally add  $(L-1)p_0$ predictors. The only difference is that the Noise-regularization adds i.i.d.~independent noise variables whereas the VAR$(L)$ adds additional  lagged  vectors.

Analogous   to the symmetry    of the AR model, the VAR model also exhibits a symmetry  property on its aggregated eigenvalues of the predictor matrix. To see this, let $X_L$ denote the $(T-L)\times NT$  predictor matrix of VAR$(L)$, and 
$EigS(L)$ denote  the sum of eigenvalues of $X_L'X_L $. 
We have the aggregated eigenvalue symmetry  for VAR$(p)$ as follows: 
$$
EigS(L) =EigS(T-L).\quad \footnote{To see this, let $\mathbb Y_{a:b}=[Y_{a},...,Y_{b}]$ whose columns are $Y_t$ ranging from $t=a,...,b$. Then  $X_L=[\mathbb Y_{L:T-1}',...,\mathbb Y_{1:T-L}']$, and $X_{T-L}$ is   rearranging elements of $X_L$, thus they have the same Frobenius norm: $\|X_L\|_F=\|X_{T-L}\|_F.$ So $EigS(L)=\tr(X_LX_L')=\|X_L\|_F^2=\|X_{T-L}\|_F^2=Eig(T-L)$.}
$$
 Thus the aggregated eigenvalues of $X_L$ do not monotonically increase as more lags are added. In contrast, the aggregated eigenvalues of the Noise-regularized regressors linearly increase as more noise are added. The sharp contrast on the aggregated eigenvalues yields a critical implication on the following quantity:
 $$
 \tr(B_IB_I'),\quad B_I= X_I(X_IX_I'+ X_NX_N')^{-1}.
 $$ 
For both models $X_I$ is the matrix of the first lag $Y_{t-1}$;     but $X_N$ corresponds to   additional lags in the VAR model and to the added noise in the Noise-regularized model. The above trace quantity  plays a pivotal role in the forecast variance: smaller $ \tr(B_IB_I')$ yields a smaller variance.

\begin{figure}[h!]
\centering
\includegraphics[width=1.0\textwidth]{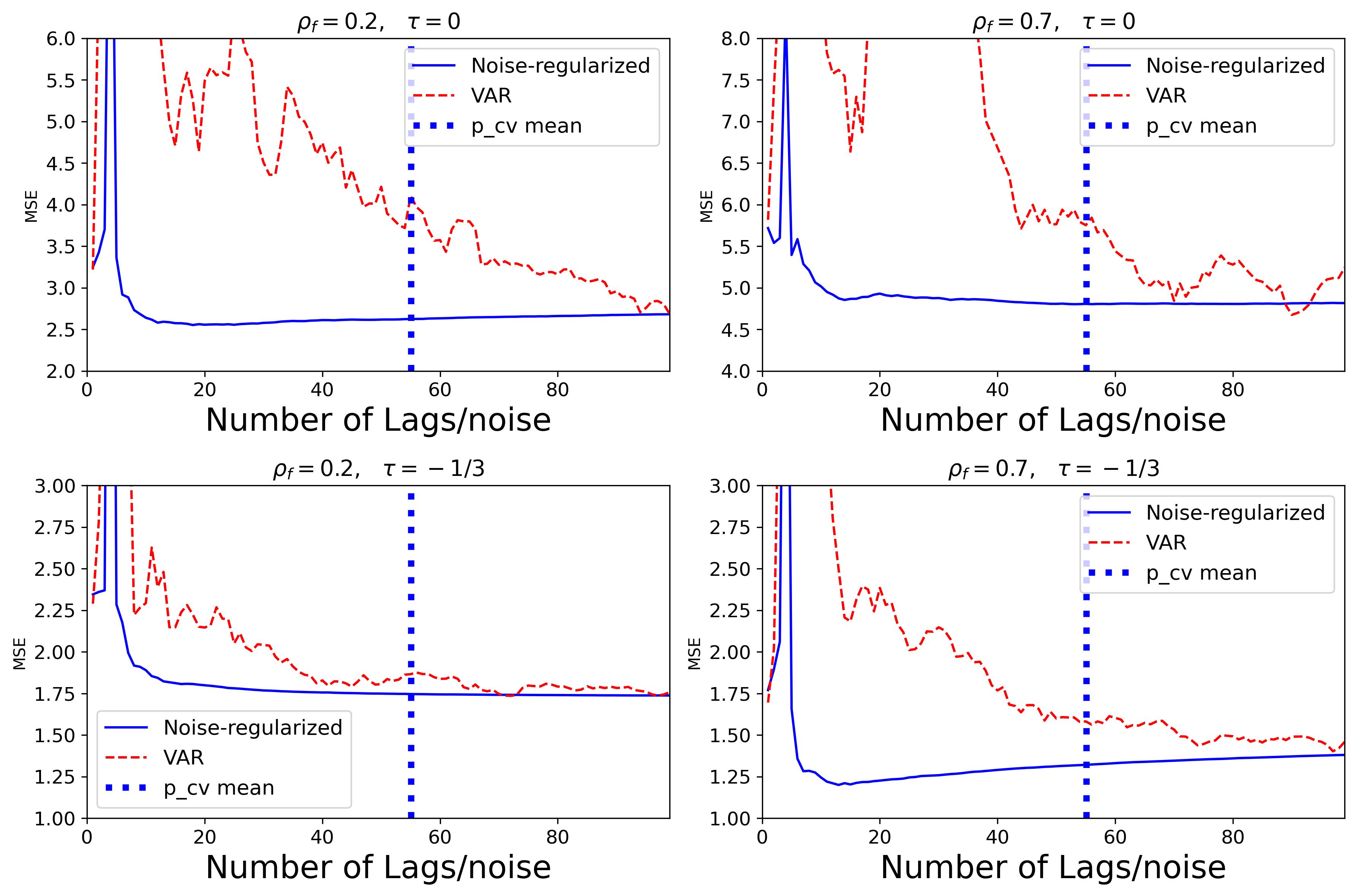}
  \caption{Comparison with VAR}
  \begin{flushleft}
\footnotesize
Notes: Predictive MSE averaged from 50 replications as the number of predictors $p$ increases. The sample size is fixed $n=100$. The Noise-regularized method uses the first $p_0$ dimensional lagged  vector as the informative predictors, and additionally adds    $(L-1)\times p_0$ noise, whereas VAR$(L)$ adds lags up to $Y_{t-L}$.    The vertical blue dashed line $p_{cv}$ in the last panel indicates the averaged $L$ chosen by the leave-one-out cross validation.
\end{flushleft}
\label{fig:var} 
\end{figure}

Figure \ref{fig:var} plots the MSE of the compared methods. In all cases, the MSE of the two methods converge as the number of added  lags/noise increase to over 80. Yet,  the cross-validation method chooses the optimal number of added noise $p_{cv}$, at which point the e Noise-regularized forecast outperforms the VAR. The comparison is particularly interesting that the comparison is qualitatively the same under both low serial dependence ($\rho_f= 0.2$) and strong serial dependence ($\rho_f=0.7$). This seems counterintuitive as the VAR model should outperform the Noise-regularized forecast under stronger serial dependence because the lagged variables carry more predictive information than pure noise. This, however, is not the case as shown in Figure \ref{fig:var}. We find two interpretations for this: First,  for the VAR model the sample size $n= T-L$ decreases as more lags are added, whereas the Noise-regularization keeps the same sample size $n= T-1$. 

\begin{figure}[h!]
%	\hspace{4em} 
\includegraphics[width=1.0\textwidth]{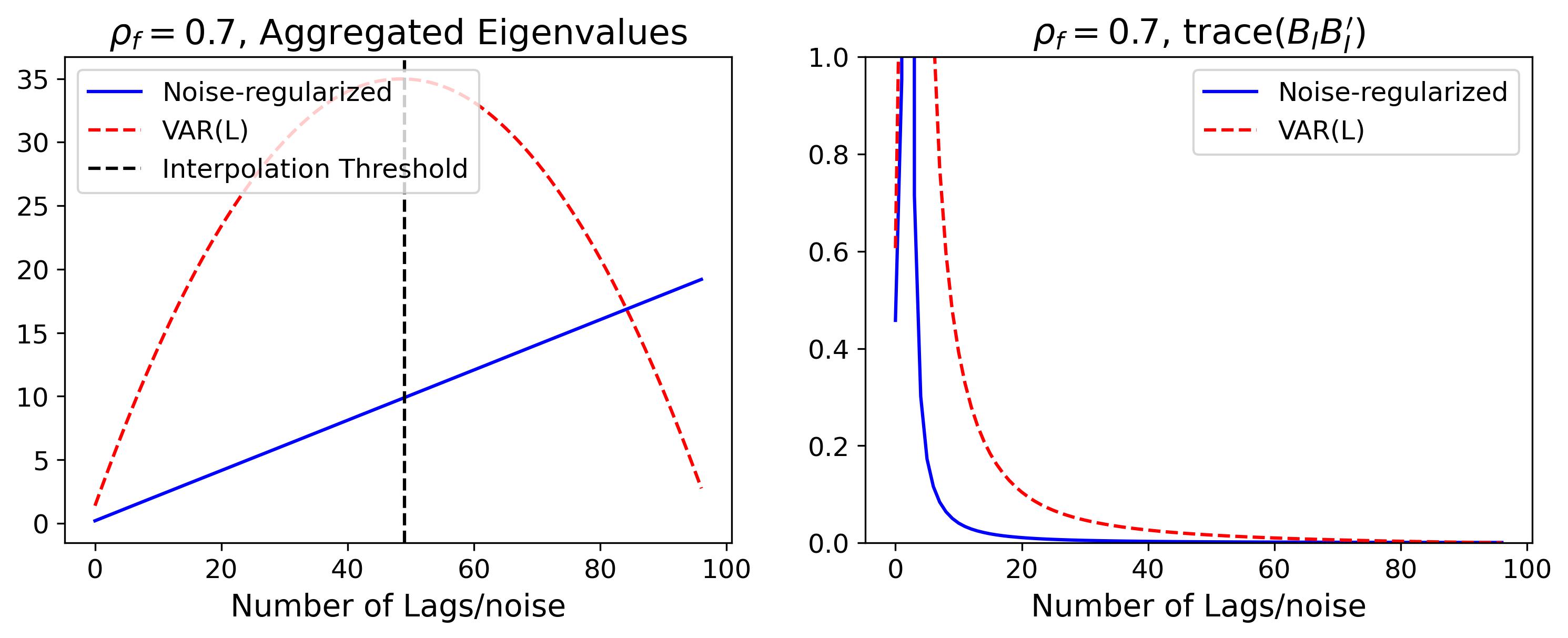}
		\caption{Aggregated Eigenvalues Comparison with VAR} \label{fig:varsymm}
\begin{flushleft}
    \footnotesize Notes: the left panel plots the sum of eigenvalues of $X_N'X_N$ where $X_N $ is the predictor matrix for VAR$(L)$ and is pure noise for the Noise-regularization. The right panel plots $\tr(B_IB_I')$ for the two models.
\end{flushleft}
\end{figure}

Secondly and more fundamentally, the symmetry  of the aggregated eigenvalues makes the VAR not benefit from regularizing eigenvalues as the Noise-regularization does.
 Figure \ref{fig:varsymm} contrasts $EigS$ and  $\tr(B_IB_I')$ for the two models in the case $\rho_f= 0.7$. It clearly shows that EigS is symmetric for   VAR  and is monotonically increasing for Noise-regularization, and that $ \tr(B_IB_I')$ of VAR   decreases slower in the over-parametrization regime.

\end{document}